\documentclass[reprint,aps,superscriptaddress,prx,longbibliography,letterpaper,amsmath,final]{revtex4-2}

\usepackage{multirow}%
\usepackage{graphicx}
\makeatletter
\newcommand{\namerefset}[1]{%
  \edef\@currentlabelname{#1}%
}
\makeatother

\usepackage{quantikz}
\usepackage{graphicx}
\usepackage{dcolumn}
\usepackage{bm}
\usepackage[colorlinks = true, linkcolor = blue, urlcolor  = blue, citecolor = blue, anchorcolor = blue]{hyperref}
\usepackage[english]{babel}
\usepackage{braket}
\usepackage{amssymb,dsfont}
\usepackage{makecell}
\usepackage{times}

\begin{document}

\title{Quantum Zeno Monte Carlo for computing observables}

\author{Mancheon Han}
\email{mchan@kias.re.kr}
\affiliation{School of Computational Sciences, Korea Institute for Advanced Study (KIAS), Seoul, 02455, Korea}
\author{Hyowon Park}
\email{hyowon@uic.edu}
\affiliation{Materials Science Division, Argonne National Laboratory, Argonne, IL, 60439, USA}
\affiliation{Department of Physics, University of Illinois at Chicago, Chicago, IL, 60607, USA}
\author{Sangkook Choi}
\email{sangkookchoi@kias.re.kr}
\affiliation{School of Computational Sciences, Korea Institute for Advanced Study (KIAS), Seoul, 02455, Korea}
\date{\today}

\begin{abstract}
  The recent development of logical quantum processors marks a pivotal transition from the noisy intermediate-scale quantum (NISQ) era to the fault-tolerant quantum computing (FTQC) era. These devices have the potential to address classically challenging problems with polynomial computational time using quantum properties. However, they remain susceptible to noise, necessitating noise resilient algorithms. We introduce Quantum Zeno Monte Carlo (QZMC), a classical-quantum hybrid algorithm that demonstrates resilience to device noise and Trotter errors while showing polynomial computational cost
  for a gapped system.
  QZMC computes static and dynamic properties without requiring
  initial state overlap or variational parameters,
  offering reduced quantum circuit depth. 
\end{abstract}

\maketitle


The quantum computer~\cite{Benioff1980,Feynman1982,NielsenAndChuang2010}
utilizes quantum algorithms to tackle computationally challenging problems,
offering potential solutions to classically hard problems.
A significant challenge lies in finding Hamiltonian eigenstates and their physical properties~\cite{Shen2023},
crucial for material design and quantum machine learning implementation.
By providing an initial state sufficiently close to the target eigenstate,
this problem can be solved within polynomial quantum time~\cite{Kitaev1995,Abrams1999}
with a fully fault-tolerant quantum computer (FTQC)~\cite{Shor1996,Gottesman1998}.
However, the preceding decades have been marked by the
noisy intermediate-scale quantum (NISQ) era~\cite{Preskill2018} rather than the FTQC era.
Due to substantial device noise, quantum algorithms for NISQ systems prioritize noise resilience,
leading to the dominance of ansatz-based algorithms~\cite{Peruzzo2014, Mcclean2016} without provable polynomial complexity.

The emergence of quantum devices with 48 logical qubits~\cite{Bluvstein2024}
marks the start of error-corrected quantum computing.
These devices, along with their future advancements,
have the potential to showcase quantum advantage, bridging the gap between NISQ and FTQC eras. Early error-corrected quantum computers are expected to handle longer quantum circuits than NISQ devices
and execute quantum algorithms with polynomial
complexity. However, algorithms designed for the
FTQC era may not be suitable for early error-corrected quantum computers,
as they still face device noise due to limited error corrections.
As a result, developing new quantum algorithms
that costs polynomial quantum time and
are resilient to noise shows promise for
achieving quantum advantage in early error-corrected quantum computers.

We introduce the quantum Zeno Monte Carlo (QZMC) algorithm.
This algorithm is robust against device noise as well as Trotter error.
Furthermore, this algorithm enables the computation of
static as well as dynamic physical properties for gapped quantum systems within
polynomial quantum time. Notably, QZMC does not necessitate overlap between the initial state and the target state, nor does it requires variational parameters.
We validate its resilience to
device noise by implementing it on IBM's NISQ devices for systems with up to 12 qubits.
We also demonstrate its resilience to the Trotter error
and the polynomial dependence of its computational cost 
by numerical demonstration on a noiseless quantum computer simulator.
Our method's resilience to Trotter errors allows us to compute eigenstate properties with shallower circuits,
as demonstrated in comparisons with recent phase estimation techniques~\cite{LinLin2022,Ding2023}.


\section*{Results}
The Quantum Zeno Monte Carlo algorithm draws 
inspiration from the quantum Zeno effect~\cite{Misra1977}. This is the phenomenon that repeated measurements slow down
state transitions. We briefly outline this effect:
A system varying with a continuous variable
$\lambda$ is represented by the state
$\ket{\psi_{\lambda}}$.
Increasing $\lambda$ to $\lambda+\Delta\lambda$
yields the state $\ket{\psi_{\lambda+\Delta\lambda}}$,
which remains $\ket{\psi_\lambda}$
with a probability of
$|\braket{\psi_\lambda|\psi_{\lambda+\Delta \lambda}}|^2$.
Because its maximum is at $\Delta\lambda=0$,
this probability becomes $1-\mathcal{O}((\Delta\lambda)^2)$
for sufficiently small $\Delta\lambda$.
By dividing $\Delta\lambda$ into $N$ slices
and measuring at each interval of $\Delta\lambda/N$,
the probability of measuring $\ket{\psi_\lambda}$
is $1 - \mathcal{O}((\Delta\lambda)^2/N)$.
Increasing the measurement frequency $N$
ensures the system remains in its initial
state $\ket{\psi_\lambda}$.

While the original article~\cite{Misra1977}
focused on state freezing
through continuous measurements,
the principle can also be applied to obtain an energy
eigenstate by varying the Hamiltonian for
each measurement~\cite{Somma2008,Poulin2009,Boixo2009,Chiang2014,LinLin2020}.
Let's denote our target Hamiltonian as $H$,
with its eigenstate as $\ket{\Phi}$.
Suppose we have an easily preparable eigenstate $\ket{\Phi_0}$ of $H_0$ and the state is adiabatically connected
to $\ket{\Phi}$. Due to the Van Vleck catastrophe~\cite{VanVleck1936,Kohn1999},
$\ket{\Phi_0}$ has very small overlap with $\ket{\Phi}$ in general,
potentially requiring a large number of measurements
to obtain $\ket{\Phi}$ directly from $\ket{\Phi_0}$.
Instead,
we consider measuring
$H_{\alpha} = (1-\lambda_\alpha) H_0 + \lambda_\alpha H$
consecutively for $\lambda_\alpha=1/N_\alpha,2/N_\alpha\dots,1$.
Utilizing the quantum Zeno principle,
we can obtain $\ket{\Phi}$ with very high probability
as we increase the number of consecutive measurements $N_{\alpha}$.

\subsection*{Quantum Zeno Monte Carlo}
\begin{figure*}[p] 
\centering
\centerline{\includegraphics[width=17.2cm]{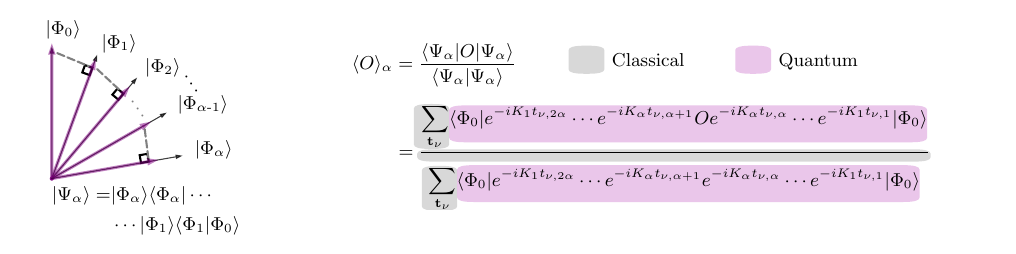}}
\caption {\textbf{Overview of the Quantum Zeno Monte Carlo}.
The construction of the unnormalized
eigenstate $\ket{\Psi_\alpha}$ of $H_\alpha$
from the eigenstate $\ket{\Phi_0}$ of $H_0$
is depicted (left). Each $\ket{\Phi_k}$ represents
the normalized eigenstate of $H_k$.
In the right, we present a summary of our
Quantum Zeno Monte Carlo for computing
the expectation value of an observable ($O$).
First, classical computer generates a time vector $\mathbf{t}_\nu = [t_{\nu,1} \,\, t_{\nu,2} \, \cdots \, t_{\nu,2\alpha}]^T$, where $t_{\nu,k}$ follows Gaussian distribution.
Next, quantum computer measure the expectation value with the given time vector.
Finally, the sum over $N_\nu$ Monte Carlo sampling as well as the division is conducted by using classical computer.
Here, $K_{\alpha'}$ represents $H_{\alpha'}-E_{\alpha'}$.
}

\label{fig:1}
\end{figure*}

\namerefset{Quantum Zeno Monte Carlo}
\label{sec:QZMC}
The quantum Zeno principle can be implemented
using projections, which is equivalent to measurements.
Let's consider 
$H_{\alpha} = (1-\lambda_\alpha) H_0 + \lambda_\alpha H$ with
$\lambda_\alpha=1/N_\alpha,2/N_\alpha\dots,1$, and $\ket{\Phi_0}$
is the eigenstate of $H_0$ that can be readily prepared.
For the eigenstate $\ket{\Phi_\alpha}$ of $H_\alpha$,
the operator that projects onto $\ket{\Phi_\alpha}$
is represented as $\ket{\Phi_\alpha}\bra{\Phi_\alpha}$.
Then, the consecutive projections $\mathcal{P}_\alpha$
applied to $\ket{\Phi_0}$ is
\begin{align}
 \ket{\Psi_\alpha} = \mathcal{P}_\alpha \ket{\Phi_0}, \quad
  \mathcal{P}_{\alpha} =\ket{\Phi_\alpha}\bra{\Phi_\alpha}\dots \ket{\Phi_1}\bra{\Phi_1},
  \label{eq:conproj}
\end{align}
which is equal to $\ket{\Phi_\alpha}$ apart from the normalization.
The quantum Zeno principle ensures that
$\braket{\Psi_{\alpha}|\Psi_\alpha}$ approaches $1$ as $N_\alpha\rightarrow\infty$.
Direct implementation of $\ket{\Phi_\alpha}\bra{\Phi_\alpha}$ is not straightforward,
and approximating it requires knowledge of the exact eigenstate, which is unknown.
To address this, we consider the projection onto
the subspace with the energy $E$.
This projection is defined as $P_H(E) = \sum_{j}\ket{j}\bra{j} \mathds{1}_{\mathcal{E}_j=E}$,
where $\mathcal{E}_j$ and $\ket{j}$ are the energy eigenvalues and eigenstates of
the Hamiltonian $H$. The function $\mathds{1}_{a=b}$ is an indicator function
that equals $1$ if $a=b$, and $0$ otherwise.
By approximating the indicator function $\mathds{1}_{\mathcal{E}_j=E}$
with the Gaussian function $\exp({-\beta^2(\mathcal{E}_j-E)^2/2})$, we
obtain the approximate projection function:
\begin{align}
P^\beta_H(E) = \sum_{j} \ket{j}\bra{j} e^{-{\beta^2(\mathcal{E}_j-E)^2}/{2}} = e^{-{\beta^2(H-E)^2}/{2}}, \label{eq:gaussproj}
\end{align}
which satisfies $\lim_{\beta\rightarrow\infty} P^\beta_H(E) = P_H(E)$.
This non-unitary operator can not be directly implemented in the quantum computer,
which only allows the unitary operation.
Instead, we use a fourier expansion~\cite{Zeng2021,Huo2023,Wang2023,Sun2023} of the approximate projection,
\begin{align}
    P^{\beta}_{H}(E) = \frac{1}{\sqrt{2\pi\beta^2}}\int_{-\infty}^{\infty} e^{-\frac{t^2}{2\beta^2}} e^{-i(H-E)t} dt. \label{eq:fourierproj}
\end{align}
Here, the integrand corresponds to Hamiltonian time evolution,
which can be simulated in polynomial time on a quantum computer~\cite{Lloyd1996,Zalka1998}.
Then, the consecutive projection $\mathcal{P}_\alpha$ can be
approximated as
\begin{align}
    \mathcal{P}^{\beta}_{\alpha} = P^{\beta}_{H_\alpha}(E_\alpha) P^{\beta}_{H_{\alpha-1}}(E_{\alpha-1}) \dots P^{\beta}_{H_{1}}(E_1), \label{eq:conproj_approx}
\end{align}
where $E_\alpha$ is the energy eigenvalue of $H_\alpha$ corresponding to
$\ket{\Phi_\alpha}$. By substituting $\mathcal{P}_\alpha$ with
$\mathcal{P}^{\beta}_{\alpha}$, the consecutive projection transforms
into a multidimensional integral of consecutive time evolution.
Using this expansion, we focus on computing the expectation values $\braket{O}$
of observables similar to recently proposed algorithms~\cite{Zeng2021,Huo2023}.
Specifically,
$\braket{O}_\alpha=\braket{\Phi_\alpha|O|\Phi_\alpha}$ is determined as
\begin{align}
  \braket{O}_\alpha = \frac{\braket{\Psi_\alpha|O|\Psi_\alpha}}{\braket{\Psi_\alpha|\Psi_\alpha}}, \label{eq:obs}
\end{align}
which requires the computation of $\braket{\Psi_\alpha|O|\Psi_\alpha}$
and $\braket{\Psi_\alpha|\Psi_\alpha}$.
For an operator $A$, $\braket{\Psi_\alpha|A|\Psi_\alpha}$
can be calculated by using approximating consecutive
projection $\mathcal{P}_\alpha$
by Eq.~(\ref{eq:conproj_approx}).
This leads to $\braket{\Psi_\alpha|A|\Psi_\alpha}\approx
\braket{\Psi^\beta_\alpha|A|\Psi^\beta_\alpha}$, where
\begin{align}
\braket{\Psi^\beta_\alpha|A|\Psi^\beta_\alpha}=\frac{1}{(2 \pi \beta^2)^{\alpha}}
\int dt_1 dt_2 \cdots dt_{2\alpha} e^{-\frac{t_1^2+t_2^2+\dots + t_{2\alpha}^2}{2\beta^2}}
\nonumber \\
\langle\Phi_0|e^{-iK_1 t_{2\alpha}} e^{-iK_2 t_{2\alpha-1}} \cdots  e^{-iK_{\alpha} t_{\alpha+1}} \nonumber \\
A e^{-iK_{\alpha} t_{\alpha}}e^{-iK_{\alpha-1} t_{\alpha-1}} \cdots e^{-iK_{1} t_{1}}|\Phi_0\rangle. \label{eq:obs2}
\end{align}
where $K_{\alpha'} $ is equal to  $H_{\alpha'}-E_{\alpha'}$ for $\alpha'=1,2,\dots,\alpha$.
This integral can be evaluated using the Monte Carlo method~\cite{Kroese2011} by sampling
$t_1, t_2, \dots t_{2\alpha}$ from a Gaussian distribution. More precisely,
\begin{align}
\braket{\Psi^\beta_\alpha|A|\Psi^\beta_\alpha}=\frac{1}{N_\nu}\sum_{\mathbf{t}_\nu} \langle\Phi_0| 
e^{-i K_1 t_{\nu,2\alpha}}
e^{-i K_2 t_{\nu,2\alpha-1}}
\cdots  \nonumber \\
e^{-i K_\alpha t_{\nu,\alpha+1}}
A e^{-i K_\alpha t_{\nu,\alpha}} e^{-i K_{\alpha-1} t_{\nu,{\alpha-1}}}\cdots
e^{-i K_1 t_{\nu,1}}|{\Phi_0}\rangle,\label{eq:obsmc}
\end{align}
where $N_\nu$ is the number of samples of 
$\mathbf{t}_\nu = [t_{\nu,1} \,\, t_{\nu,2} \, \cdots \, t_{\nu,2\alpha}]^T$.
Each $t_{\nu,k}$ is drawn from a Gaussian distribution with a standard deviation of $\beta$.
We refer to this approach as the quantum Zeno Monte Carlo (QZMC) method.
From its formulation, it is evident that QZMC can be used to compute various
static and dynamic properties of Hamiltonian eigenstates. Figure~\ref{fig:1}
provides a summary of the method.

Among various eigenstate properties, the energy eigenvalue holds prime importance,
as it is essential for QZMC to perform the approximate projection $P_H(E)$.
In this section, we describe the method for computing energy eigenvalues
using Quantum Zeno Monte Carlo.
QZMC employs eigenstates $\ket{\Psi_\alpha}$, which satisfy
$\langle \Psi_{\alpha}|\Phi_\alpha\rangle\langle\Phi_\alpha|(H_{\alpha}-H_{\alpha-1})|\Psi_{\alpha-1}\rangle
= (E_{\alpha}-E_{\alpha-1}) \langle \Psi_{\alpha} | \Psi_{\alpha} \rangle$.
Using this, the energy eigenvalue is estimated from
\begin{align}
    E_{\alpha} = E_{\alpha-1} + \frac{\braket{\Psi_\alpha\ket{\Phi_\alpha}{\bra{\Phi_\alpha}}(H_\alpha-H_{\alpha-1})|\Psi_{\alpha-1}}}{\braket{\Psi_\alpha|\Psi_\alpha}}.\label{eq:corrector}
\end{align}
This equation can be computed using the same strategy we used in Eq.~(\ref{eq:obsmc}).
Compared to estimating entire energy from $\braket{H_\alpha}_\alpha$ using Eq.~(\ref{eq:obs}), 
this approach improves robustness against noise
by limiting its impact to the energy difference alone.
Building on this insight, we propose the predictor-corrector QZMC
method for determining energy eigenvalues. Suppose
we know $E_0, E_1, \dots, E_{\alpha-1}$ and seek to compute $E_{\alpha}$.
Inspired by the predictor-corrector method commonly used for solving
differential equations~\cite{HeathSC}, we begin with an initial estimate of $E_{\alpha}$,
referred to as the predictor. Various approaches can be employed to determine
the predictor. One frequently used method in this manuscript is
the first-order perturbation approximation~\cite{LandauQM}, given by
$E_{\alpha} = E_{\alpha-1} + \langle \Phi_{\alpha-1} | (H_{\alpha}-H_{\alpha-1}) | \Phi_{\alpha-1} \rangle$.
Here, $\langle \Phi_{\alpha-1} | (H_{\alpha}-H_{\alpha-1}) | \Phi_{\alpha-1} \rangle$
is computed using Eq.~(\ref{eq:obs}).
Using the predictor $E_{\alpha}$,
we then compute a more accurate estimate of $E_{\alpha}$ using Eq.~(\ref{eq:corrector}).
Further details of the QZMC method, including formulations for the computation
of Green's functions, are provided in the Supplementary Information Sec.~\uppercase{i}.

In the formulation of the method, we began with $H_0$, which can be easily solved on
a classical computer, and whose eigenstate $\ket{\Phi_0}$ is readily preparable as
a quantum circuit.
Notably, $\ket{\Phi_0}$ is not required to have
a finite overlap with the target eigenstate $\ket{\Phi}$.
However, the synthesis of arbitrary unitary operations can
incur exponential quantum time costs~\cite{Shende2006}, making the preparation
of $\ket{\Phi_0}$ challenging even when $H_0$ is exactly solvable on a classical
computer.
Our method can also be applied in such cases by following an alternative procedure.
First, we prepare an easily accessible state $\ket{\tilde{\Phi}_0}$ with
a finite overlap with $\ket{\Phi_0}$ (e.g., $|\braket{\Phi_0|\tilde{\Phi}_0}|^2 > 0.5$).
Then, we project $\ket{\tilde{\Phi}_0}$ onto $\ket{\Phi_0}$ using Eq.~(\ref{eq:gaussproj}) and perform QZMC in an equivalent way. Consequently,
\begin{align}
\ket{\Psi_\alpha} = \mathcal{P}_\alpha \ket{\Phi_0} \braket{\Phi_0|\tilde{\Phi}_0}, \label{eq:state_prep}
\end{align}
is used instead of Eq.~(\ref{eq:conproj}).
As $\ket{\Phi_0}$ is known and can be processed on a classical computer, finding $\ket{\tilde{\Phi}_0}$ can be efficiently accomplished using classical computing resources. Thus, applying QZMC is feasible even for systems where $\ket{\Phi_0}$ is not easily preparable

Finally, we note that the transformation in Eq.~(\ref{eq:fourierproj})
can be interpreted as the Hubbard-Stratonovich
transformation~\cite{Hubbard1959,Stratonovich1958,Zhang2013},
which underpins the auxiliary-field quantum Monte Carlo (AFQMC)
method~\cite{Zhang2003,Motta2018}. AFQMC is a widely-used classical
approach for computing ground state properties of quantum many-body systems.
In AFQMC, the Hubbard-Stratonovich transformation is employed to transform two-body
interactions term into one-body term at the cost of introducing auxiliary fields.
In contrast, QZMC leverages a similar transformation to express non-unitary operators
as integrals over unitary operations, enabling its implementation on quantum computers.
Unlike AFQMC or diffusion Monte Carlo (DMC)~\cite{Ceperly1986},
which iteratively adjust the trial energy as random walkers
propagate in imaginary time under a fixed Hamiltonian,
QZMC calculates the ground-state energy by integrating
the energy difference formula (Eq.~(\ref{eq:corrector})) while gradually changing
the Hamiltonian toward the target Hamiltonian.

\subsection*{Error analysis and Cost estimation}
This section provides an error analysis and cost estimation
for our method. A detailed analysis is available in Sec.~\uppercase{ii} of the Supplementary Information.
For simplicity, we assume a linear interpolation between $H_0$ and $H$,
defined as $H_\alpha = H_0 + \lambda_\alpha H'$, where $H' = H - H_0$
and $\lambda_\alpha = 1/N_\alpha, 2/N_\alpha, \dots, 1$. We
also assume the target state is gapped from other states, with a lower bound $\Delta_g$
on the energy gap.
The computational cost is evaluated in terms of circuit depth 
and the number of circuits ($N_\nu$) required. Circuit depth depends on
$N_\alpha$ and systematic errors from $\beta$,
while the number of circuits accounts for statistical errors arising from Gaussian sampling
of $t_\nu$.
The goal is to estimate the energy eigenvalue 
within an error $\epsilon$.
From the formulation (e.g., Eq.~(\ref{eq:obs}), (\ref{eq:corrector})),
it is essential to maintain a finite value of $\braket{\Psi^\beta_\alpha|\Psi^\beta_\alpha}$
for a feasible computation. We first analyze error
of $\braket{\Psi_\alpha^\beta|\Psi_\alpha^\beta}$ and address
the condition under which
$\braket{\Psi^\beta_\alpha|\Psi^\beta_\alpha} \geq (1 - \eta)$
for $\eta \in (0,1)$.

\subsection*{Error analysis of $\braket{\Psi^\beta_\alpha|\Psi^\beta_\alpha}$}
\namerefset{Error analysis of $\braket{\Psi^\beta_\alpha|\Psi^\beta_\alpha}$}
\label{sec:error_analysis_eta}
Our analysis begins with the assumption of exact projection. We then incorporate the effects of finite $\beta$,
trotterization, and $N_\nu$.
We decompose $\eta$ as $\eta_0 + \delta\eta_\beta + \delta\eta_T + \delta \eta_{\textrm{mc}} $,
where $\eta_0$ corresponds to exact projection,
$\delta\eta_\beta$ represents the error due to finite $\beta$,
$\delta\eta_T$ arises from trotterization,
and $\delta\eta_{\textrm{mc}}$ reflects the finite number of samplings.

First, under the assumption of exact projection,
we estimate the number of projections $N_\alpha$
required to satisfy $\braket{\Psi_\alpha|\Psi_\alpha} \geq 1-\eta_0$.
Applying perturbation theory~\cite{LandauQM}, we obtain
\begin{align}
|\braket{\Phi_\alpha|\Phi_{\alpha+1}}|^2 \geq 1-\|H'\|^{2}\Delta_g^{-2}N_\alpha^{-2} \label{eq:proj1}
\end{align}
up to the leading order in $N_\alpha^{-1}$.
Consequently, $\braket{\Psi_\alpha|\Psi_\alpha} = |\braket{\Phi_0|\Phi_{1}}|^2
|\braket{\Phi_1|\Phi_{2}}|^2 \cdots|\braket{\Phi_{\alpha-1}|\Phi_{\alpha}}|^2$
is bounded below by $1-\|H'\|^2/\Delta_g^{-2} N_\alpha^{-1}$. By setting $N_\alpha \geq \|H'\|^{2}\Delta_g^{-2}\eta_0^{-1}$,
we ensure that $\braket{\Psi_\alpha|\Psi_\alpha}\geq 1-\eta_0$.
For the ground state, a smaller $N_\alpha$
can be used due to the ground state property,
yielding \begin{align}
N_\alpha\geq \|H'\|\Delta_g^{-1}\eta_0^{-1}. \label{eq:proj2}
\end{align}
Please see Sec.~\uppercase{ii}\,A\,1 of Supplementary Information for derivations
of Eq.~(\ref{eq:proj1}) and Eq.~(\ref{eq:proj2}).

Next, we examine the effect of finite $\beta$.
The error in the projected state due to finite $\beta$ can be written as
$\ket{\delta\Psi_\alpha^\beta} = \ket{\Psi_\alpha^\beta}-\ket{\Psi_\alpha}$.
Perturbative analysis shows that
\begin{align}
  \|\ket{\delta\Psi_\alpha^\beta}\|\leq  (\alpha/N_\alpha) e^{-\beta^2\Delta_g^2/2} \|H'\|\Delta_g^{-1}, \label{eq:betabound}
\end{align}
up to the leading order in $1/N_\alpha$.
As a result, $\delta\eta_\beta\leq 2 \exp(-\beta^2\Delta_g^2/2) \|H'\|\Delta_g^{-1}$.
By choosing \begin{align}
\beta\geq
\Delta_g^{-1}\sqrt{2}\operatorname{log}^{1/2}(2\|H'\|\Delta_g^{-1}(\eta-\eta_0)^{-1}),
\end{align}
we ensure that $\eta_0+\delta\eta_\beta \leq \eta$.

For time evolution, we primarily use trotterization.
The circuit depth required for our method is determined by the total number of trotterization steps.
The error in the projected state due to trotterization is expressed as
$\ket{\delta\Psi_\alpha^{\beta,T}}=\ket{\Psi_\alpha^{\beta,T}}-\ket{\Psi_\alpha^{\beta}}$,
where $\ket{\delta\Psi_\alpha^{\beta,T}}$
rises from trotterized time evolutions. The trotterization error for each
$\alpha$-th time evolution with evolution time $t$ is bounded by
$C_{\alpha,p} |t|^{1+p}N_{T,\alpha}^{-p}$, where $N_{T,\alpha}$ is the number of
trotterization steps for each $\alpha$,
$p$ is the trotterization order,
and $C_{\alpha,p}$ is
the coefficient which is proportional to the sum of the norms of the commutators~\cite{Childs2021}.
Then, we can show
\begin{align}
\|\ket{\delta\Psi_\alpha^{\beta,T}}\|\leq
\sum_{\alpha'=1}^\alpha C_{\alpha',p}M_{1+p}(\beta) N_{T,\alpha'}^{-p}, \label{eq:trotterbound}
\end{align}
and $\delta\eta_T \leq 2 \sum_{\alpha'=1}^{N_\alpha} C_{\alpha',p}M_{1+p}(\beta)  N_{T,\alpha'}^{-p}$,
up to the leading order of $N_{T,\alpha}^{-1}$.
Here $M_{1+p}(\beta)$ is the expectation value of $|t|^{1+p}$ for a
Gaussian distribution with a standard deviation of $\beta$.
To ensure the trotterization error is smaller than $\delta\eta_T$,
the total number of trotter steps $N_T = 2 \sum_{\alpha=1}^{N_\alpha} N_{T,\alpha}$
can be chosen as
\begin{align}
N_T \geq 2 \left(\frac{2N_\alpha}{\delta\eta_T}\right)^{1/p} \sum_{\alpha=1}^{N_\alpha} C_{\alpha,p}^{1/p} M^{1/p}_{1+p}(\beta) , \label{eq:trottercondition}
\end{align}
with each $N_{T,\alpha}$ proportional to $C_{\alpha,p}^{1/p} M^{1/p}_{1+p}(\beta)$.

Finally, we consider the statistical error $\delta\eta_{mc}$, which arises
arises from the finite number of samples $N_\nu$.
Defining $x(\mathbf{t})=
\langle\Phi_0| e^{-iK_1 t_{2\alpha}}  e^{-iK_2 t_{2\alpha-1}} \allowbreak \cdots  e^{-iK_{\alpha} t_{\alpha+1}} \allowbreak
A e^{-iK_{\alpha} t_{\alpha}} e^{-iK_{\alpha-1} t_{\alpha-1}}\cdots e^{-iK_{1} t_{1}}|\Phi_0\rangle$,
$g(\mathbf{t})\allowbreak=(2 \pi \beta^2)^{-\alpha} e^{-(t_1^2+t_2^2+\dots + t_{2\alpha}^2)/(2\beta^2)}$,
Eq.~(\ref{eq:obs2}) can be seen as finding the
expectation value $E[x]$ of $x(\mathbf{t})$
with the probability of $g(\mathbf{t})$.
The case of $A=\mathbb{I}$ corresponds to
$\braket{\Psi^\beta_\alpha|\Psi^\beta_\alpha}$.
The variance of x, $\sigma_x^2$, is given by $E[x^2]-(E[x])^2$.
Since $\|x(\mathbf{t})\| \leq 1$, $\sigma_x^2\leq 1 - (E[x])^2$.
So, the standard error of $\braket{\Psi^\beta_\alpha|\Psi^\beta_\alpha}$
using $N_\nu$ samples is bounded by ${{N_\nu}}^{-1/2} (2\eta - \eta^2)$.
Therefore, getting $\braket{\Psi^\beta_\alpha|\Psi^\beta_\alpha}$
with desired precision $\delta\eta_\textrm{mc}$ will requires
\begin{align}
N_\nu \geq (2\eta-\eta)^2\delta\eta_\textrm{mc}^{-2}.
\end{align}
In the sampling procedure, an additional source of statistical error, known as shot noise, arises.
On currently accessible quantum computers, each circuit is measured
with $N_s$ repeated measurements, referred to as ``shots".
The finite number of shots introduces a standard error of $1/\sqrt{N_s}$
for each measurement. This modified the statistical error dependence from
${N_v}^{-1/2}$ to ${N_v}^{-1/2}(1+{N_s}^{-1/2})$.

\subsection*{Error analysis of $E_\alpha$}
The error $\epsilon$ of $E_\alpha$ is analyzed similarly to $\braket{\Psi_\alpha^\beta|\Psi_\alpha^\beta}$.
Like $\eta$, $\epsilon$ is decomposed as $\epsilon_\beta + \epsilon_T + \epsilon_\textrm{mc}$,
Here, $\epsilon_\beta$ arises from the finite $\beta$,
$\epsilon_T$ is due to trotterization, and $\epsilon_\textrm{mc}$
results from the finite number of samples.

First, we consider the energy estimation error arises from finite $\beta$, $\epsilon_\beta$.
In our method, the energy difference is computed based on Eq.~(\ref{eq:corrector}).
Each energy difference estimator introduces an error of order 
$N_\alpha^{-1} \|H'\| \|\ket{\delta\Psi_{\alpha-1}^\beta}\|/(1-\eta)$.
Detailed calculations in Sec.~\uppercase{ii}\,B\,1 of the Supplementary Information
show that the proportionality constant is $4$.
Thus $\epsilon_\beta \leq \sum_{\alpha=1}^{N_\alpha} 4 N_\alpha^{-1} \|\ket{\delta\Psi_{\alpha-1}^\beta}\| \|H'\| /(1-\eta)$. From Eq.~(\ref{eq:betabound}),
we have
\begin{align}
\epsilon_\beta \leq 2 \exp(-\beta^2\Delta_g^2/2) \|H'\|^2 \Delta_g^{-1}/(1-\eta)
\end{align}
To ensure the projection error is smaller than $\epsilon_\beta$,
we use $\beta$ that satisfy
\begin{align}
\beta\geq \Delta_g^{-1} \sqrt{2}\operatorname{log}^{1/2}(2\|H'\|^2\Delta_g^{-1}(1-\eta)^{-1}\epsilon_\beta^{-1}).
\end{align}

The discussion of the trotterization error follows a similar approach to that of $\beta$.
The error $\epsilon_T$ is bounded as $\epsilon_T \leq \sum_{\alpha=1}^{N_\alpha}
4 N_\alpha^{-1}\|\ket{\delta\Psi_{\alpha-1}^{\beta,T}}\| \|H'\|
/ (1-\eta)$.
Using Eq.~(\ref{eq:trotterbound}) and  
assuming $N_{T,\alpha}$ is determine to be proportional
to $C_{\alpha,p}^{1/p} M^{1/p}_{1+p}(\beta)$, the error can be expressed as
\begin{align}
\epsilon_T \leq 2 N_\alpha \left(\frac{N_T}{2}\right)^{-p}  \left(\sum_{\alpha=1}^{N_\alpha} C_{\alpha,p}^{1/p} M^{1/p}_{1+p}(\beta)\right)^p \frac{\|H'\|}{1-\eta} .
\end{align}
To achieve a desired $\epsilon_T$, the total number of trotter steps can be chosen
as
\begin{align}
  N_T \geq 2 \frac{(2 N_\alpha \|H'\|)^{1/p}}{(\epsilon_T (1-\eta))^{1/p}}\sum_{\alpha=1}^{N_\alpha} C_{\alpha,p}^{1/p} M^{1/p}_{1+p}(\beta).
\end{align}
In practice, Trotter errors are considerably smaller
than the theoretical bounds~\cite{Childs2021,Layden2022}.
Additionally, as discussed in \nameref{sec:noise_resilience} section,
error cancellation occurs between the numerator and the denominator.
Consequently, the number of Trotter steps required
is substantially lower than the theoretical estimate.

To estimate the statistical error $\epsilon_\textrm{mc}$ in the
energy calculation, we examine Eq.~(\ref{eq:corrector}).
The numerator in this equation is computed through a Monte Carlo summation
of $\langle\Phi_0|e^{-i K_1 t_{\nu,2\alpha}} \allowbreak
e^{-i K_2 t_{\nu,2\alpha-1}}
\cdots 
e^{-i K_\alpha t_{\nu,\alpha+1}} 
e^{-i K_\alpha t_{\nu,\alpha}} \Delta\lambda H'
e^{-i K_{\alpha-1} t_{\nu,\alpha-1}}
\allowbreak
e^{-i K_{\alpha-2} t_{\nu,\alpha-2}}
\cdots
e^{-i K_1 t_{\nu,1}}|{\Phi_0}\rangle$.
Because time evolutions are unitary, each term in the summation is bounded by $\Delta\lambda\|H'\|$.
This results in a Monte Carlo error of the numerator bounded by $\Delta\lambda\|H'\|/\sqrt{N_\nu}$.
Taking into account the effect of the denominator and
summing over $\alpha$ from $1$ to $N_\alpha$, we find that the total error is bounded by
\begin{align}
\epsilon_{\textrm{mc}} \leq \|H'\|/\sqrt{N_\nu} (1-\eta)^{-1}(1+(1-\eta)^2)^{-1/2}. \label{eq:energy_sampling_bound}
\end{align}
The statistical precision of $\epsilon_{\textrm{mc}}$
can be achieved by using $N_\nu$ such that
\begin{align}
N_\nu \geq \epsilon_{\textrm{mc}}^{-2}\|H'\|^2 (1-\eta)^{-2}(1+(1-\eta)^2). \label{eq:minimumsampling}
\end{align}

\subsection*{Computational cost}
Based on the error analysis discussed, we estimate the
computational cost of determining the ground state energy using
QZMC and summarize the results in Table~\ref{tab:cost}.

First, we discuss the circuit depth required to estimate ground state energy
using QZMC.
Excluding the cost of preparing the initial state, the circuit depth required for our method is
determined by the total time evolution length, which is proportional to $\beta N_\alpha$.
From the previous discussion, $N_\alpha \propto \Delta_g^{-1} \|H'\|$, so $ N_\alpha =\mathcal{O}(\Delta_g^{-1}\operatorname{poly}(n))$.
Similarly, $\beta \propto \Delta_g^{-1} (\log(2\|H'\|\Delta_g^{-1}(1-\eta)^{-1}\epsilon^{-1}))^{1/2}$,
$\beta = \mathcal{O}(\Delta_g^{-1}\log^{1/2}(\Delta_g^{-1}\epsilon^{-1}n))$.
Therefore, the total time evolution length required for our method is
$\mathcal{O}(\Delta_g^{-2}\operatorname{log}^{1/2}(\Delta_g^{-1}\epsilon^{-1}n)\operatorname{poly}(n))$.

The practical implementation of our method requires trotterization,
so the circuit depth for QZMC is determined by the total number of Trotter steps $N_T$.
From the previous discussion,
$N_T \propto \epsilon^{-1/p} \|H'\|^{1/p} N^{1/p}_\alpha (\sum_\alpha C_{\alpha,p}^{1/p} M^{1/p}_{1+p}(\beta)$,
where $p$ is the order of trotterization.
Since $C_{\alpha,p} = \mathcal{O}(\operatorname{poly}(n))$~\cite{Childs2021} and
$M^{1/p}_{1+p}(\beta) = \mathcal{O}(\beta^{(1+1/p)})$,
$N_T = \mathcal{O}(\epsilon^{-1/p}\operatorname{poly}(n)(\beta N_\alpha)^{1+1/p})$.
Substituting $\beta$ and $N_\alpha$, we have
\begin{align}
N_T = \mathcal{O}(\epsilon^{-\frac{1}{p}}\Delta_g^{-2-\frac{2}{p}}\operatorname{log}^{\frac{1}{2}+\frac{1}{2p}}(\Delta_g^{-1}\epsilon^{-1}n) \operatorname{poly}(n)).
\end{align}

Second, we discuss the total number of samples required to
estimate ground state energy within a precision of $\epsilon$.
From Eq.~(\ref{eq:minimumsampling}), the number of samples
$N_\nu$ required to achieve a precision $\epsilon$
is $\mathcal{O}(\epsilon^{-2}\operatorname{poly}(n))$.
Since QZMC should be performed for
$\alpha = 1, 2, \dots, N_\alpha$, the total number of samples required is
$\mathcal{O}(\epsilon^{-2}\operatorname{poly}(n)N_\alpha)=
\mathcal{O}(\Delta_g^{-1}\epsilon^{-2}\operatorname{poly}(n))$.

\begin{table*}[p] 
\caption{Computational cost of QZMC and other quantum algorithms}
\label{tab:cost}%
\begin{tabular}{@{}lll@{}}
\hline
         & Maximum time evolution length & Total number of samples\\
\hline
QZMC         &$\mathcal{O}(\Delta_g^{-2}(\operatorname{log}(\Delta_g^{-1}\epsilon^{-1}n))^{1/2} \operatorname{poly}(n))$ & $\mathcal{O}(\epsilon^{-2}\Delta_g^{-1}\operatorname{poly}(n))$  \\
QPE~\cite{Berry2009,Higgins2007}  &$\tilde{\mathcal{O}}(\epsilon^{-1}p_0^{-1})$ & $\tilde{\mathcal{O}}(p_0^{-1}\operatorname{polylog}(\epsilon^{-1}))$ \\
QEEA~\cite{Somma2019} &$\tilde{\mathcal{O}}(\epsilon^{-1}\operatorname{polylog}(p_0^{-1}))$  & $\tilde{\mathcal{O}}(\epsilon^{-3}p_0^{-2})$  \\
Ref.~\cite{LinLin2022}&$\tilde{\mathcal{O}}(\epsilon^{-1}\operatorname{polylog}(p_0^{-1}))$  & $\tilde{\mathcal{O}}(p_0^{-2}\operatorname{polylog}(\epsilon^{-1}))$ \\
Ref.~\cite{Wang2023}  &$\mathcal{O}(\Delta_g^{-1}\operatorname{polylog}(\epsilon^{-1}p_0^{-1}\Delta_g))$  & $\mathcal{O}(p_0^{-2}\epsilon^{-2}\Delta_g^2)$ \\
\botrule
\end{tabular}
\footnotetext{
This table summarize the cost of QZMC to compute
the ground state energy
and compares it with
several other quantum algorithms that computes
the ground state energy
within a single ancilla qubit. Complexity analysis of
QPE and QEEA imported from Ref.~\cite{LinLin2022}. Here, $p_0$ the probability of getting exact eigenstate
from the initial states, $\epsilon$ is a desired precision in the energy,
$n$ is the number of qubits,
and $\Delta_g$ is the lower bound of the energy gap between the ground and other states.
Optimized algorithms for highly overlapped
initial states~\cite{Ding2023,Ni2023} shows similar dependence with algorithm
of Ref.~\cite{LinLin2022}, only constant factor is different.
}
\end{table*}

\subsection*{Remarks}
A key characteristic of our method is that the approximate projection depends on the energy estimate
$\epsilon$,
meaning the calculational precision can affect subsequent calculations.
If $\epsilon$ comparable to or larger than $\Delta_g$, the approximate projection
fails to target the desired states, making the calculations infeasible.
For $\epsilon$ much smaller than $\Delta_g$, the projected state
becomes $\exp(-\alpha \beta^2\epsilon^2/2)\ket{\Psi_\alpha^\beta}$, inducing
attenuation of $r_a=\exp(-N_\alpha \beta^2\epsilon^2)$ of
$\braket{\Psi_\alpha^\beta|\Psi_\alpha^\beta}$ for $\alpha=N_\alpha$.
To ensure $r_a\geq r$ for some finite $r$,
$\beta$ should satisfy
$\beta\leq\epsilon^{-1}N_\alpha^{-1/2}\log^{-1/2}(1/r)$.
Thus the energy estimate precision $\epsilon$ imposes a limit on $\beta$.

Another aspect worth addressing is the potential for a sign problem.
The error analysis and computational cost estimation indicate
that our method is, in principle,
free from the sign problem for gapped systems.
For such systems, for any $\eta \in (0,1)$, there exist
sufficiently large parameters $\beta$, $N_\alpha$, and $N_\nu$, scaling polynomially
with the number of qubits $n$, such that $\braket{\Psi_\alpha^\beta|\Psi_\alpha^\beta}$
is lower-bounded by $1-\eta$.
In practice, error sources such as Trotter errors and device noise reduce
$\braket{\Psi_\alpha^\beta|\Psi_\alpha^\beta}$, resulting in
noise amplification in Eq.~(\ref{eq:obs}) and Eq.~(\ref{eq:corrector}),
analogous to the conventional sign problem in Monte Carlo methods.

The realization of our method requires computing
the overlap between the initial and time-evolved states
on a quantum computer.
In the most general setting, this involves
controlled time evolution~\cite{NielsenAndChuang2010}, which demands attaching control
lines to every gate, making it resource-intensive. However, if $H_\alpha$ shares
a common eigenstate, controlled time evolution can be avoided, as shown in other
methods~\cite{LinLin2022}. Since chemical and physical Hamiltonians often share
a common eigenstate, such as the vacuum, this feature makes our method practical
for applications in chemistry and physics. For
the specific form of the quantum circuit used in our method,
see Sec.~\uppercase{ii}\,D\,2 of the Supplementary Information.

\subsection*{Applications of QZMC}\label{sec:applications}

Here, we verify our method by applying it to solve various quantum many body
systems.

\begin{figure} 
\centering
\centerline{\includegraphics[width=8.6cm]{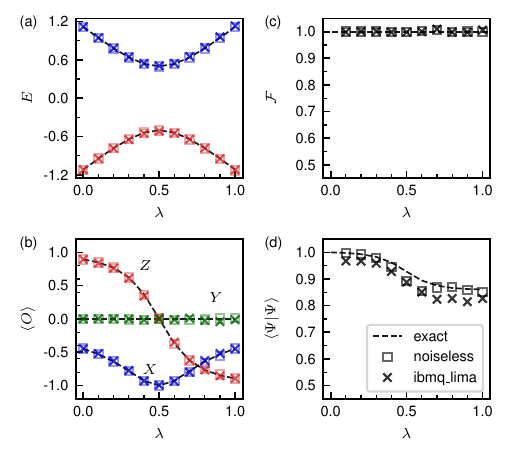}}
\caption {\textbf{A one-qubit system}.
The energy eigenvalues of the ground (red) and the excited state (blue) are plotted
in (a).
In (b), we plotted $\braket{X}$ (blue), $\braket{Y}$ (green), and
$\braket{Z}$ (red) calculated for the ground states. 
(c) and (d) display the fidelity $\mathcal{F}$
and $\braket{\Psi|\Psi}$ for the ground state. In (a)-(d),
dotted lines represent
the exact result, boxes represent QZMC results with a noiseless simulator,
and crosses represent results with
\textbf{\textit{ibmq}}\underline{ }\textbf{\textit{lima}}.
In this figure, we used $\beta=5$ and $N_\nu=400$.}
\label{fig:2}
\end{figure}

\begin{figure*} 
\centering
\centerline{\includegraphics[width=17.2cm]{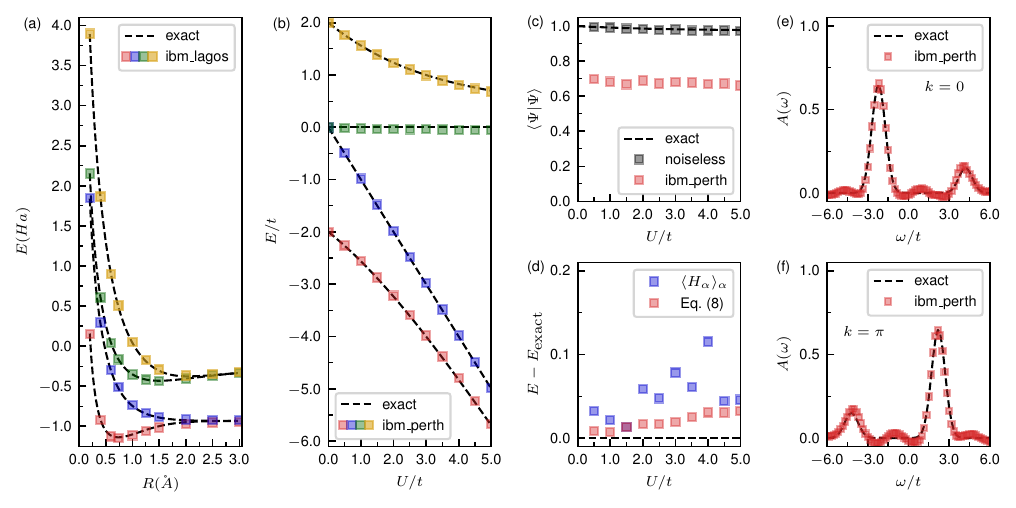}}
\caption {\textbf{H$_2$ and the Hubbard dimer}.
(a) plots energy eigenvalues of H$_2$ in a STO-3G basis
as a function of the bond length. Here, we used $\beta=5$
and NISQ device calculation is conducted with
\textbf{\textit{ibm}}\underline{ }\textbf{\textit{lagos}}.
In (b)-(f), we considered the Hubbard dimer. (b)
shows energy eigenvalues
as a function of the Coulomb interaction $U$.
In (a) and (b), different states are distinguished by different colors.
In (c), we compared $\braket{\Psi|\Psi}$ of
the ground state calculated with the NISQ device
with exact values and noiseless QZMC results.
(d) compares two energy estimator $\braket{H_\alpha}_\alpha = \braket{\Phi_\alpha| H_\alpha |\Phi_\alpha}$
and Eq.~(\ref{eq:corrector}).
The spectral functions for two different crystal momentum (e) $k=0$
and (f) $k=\pi$ are plotted.
For the Hubbard dimer, we used $\beta=0.5$ and 
\textbf{\textit{ibm}}\underline{ }\textbf{\textit{perth}}
is used. In this figure, we used 
$N_\nu=100$ Monte Carlo samples for each $\alpha$ and 
and the spectral function is calculated with $300$ Monte Carlo samples.
}

\label{fig:3}
\end{figure*}

\begin{figure} 
\centering
\centerline{\includegraphics[width=8.6cm]{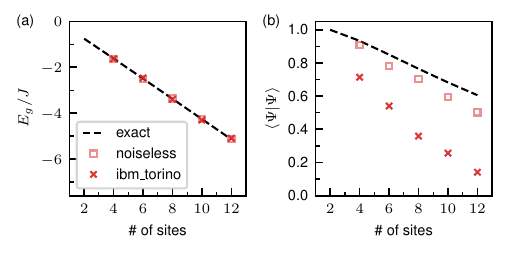}}
\caption {\textbf{NISQ simulation of XXZ model}
(a) The energy eigenvalues and (b) $\braket{\Psi|\Psi}$ of XXZ model for various sizes from $4$ to $12$ qubits
are plotted. In this figure, dotted lines
represent the calculation with exact projection, boxes represent noiselees simulation result,
and crosses are QZMC results with the \textbf{\textit{ibm}}\underline{ }\textbf{\textit{torino}}.
For QZMC, we used $\beta=\sqrt{2}$, $N_\alpha=1$, and $N_\nu=300$. 
}
\label{fig:4}
\end{figure}

First, we used our method to compute physical properties
with NISQ devices.
The first system we consider (Figure~\ref{fig:2})
is the one-qubit system with the Hamiltonian.
$H(\lambda) = X/2 + (2\lambda-1) Z$.
Next, we simulate the H$_2$ molecule (Figure~\ref{fig:3}~(a))
in the STO-3G basis~\cite{Stewart1970},
a typical testbed for quantum algorithms~\cite{O'Malley2016,Motta2020}.
By constraining the electron number
to be 2 and the total spin to be 0~\cite{Seeley2012,Steudtner2018},
the system can be represented by a 2-qubit Hamiltonian.
We calculate the energy spectrum of $4$ low-lying eigenstates of H$_2$ as a function of interatomic distance ($R$).
Then, we consider the $2$-site Hubbard model~\cite{Hubbard1963}, the Hubbard dimer.
The Hubbard dimer (Figure~\ref{fig:3}~(b)-(f)) at its half filling and singlet spin configuration can also be mapped to a two-qubit Hamiltonian.
$4$ low-lying Energy eigenvalues of the Hubbard dimer are computed by increasing onsite Coulomb
interaction($U$) from $0$. 
For these calculations, we create a discrete path with $N_\alpha=10$,
and apply the predictor-corrector QZMC for $H_\alpha=H(\lambda_\alpha)$.
Lastly, we applied our method to the XXZ model (Figure~\ref{fig:4}) in one-dimension, which has the Hamiltonian  
\begin{align}
H = -J \sum_{i=1}^{n-1} \left(S_i^x S_{i+1}^x + S_i^y S_{i+1}^y + \Delta S_i^z S_{i+1}^z\right).
\end{align}
We computed systems with $n=4$ to $n=12$, using $J=1$ and $\Delta=-1$.
For a quantum circuit implementation of trotterization for XXZ model,
we used recently suggested optimized circuit~\cite{Chowdhury2024},
with two trotter steps.

The one-qubit system results are displayed in Fig.~\ref{fig:2}.
Fig.~\ref{fig:2}~(a) shows the ground and the excited state
energy eigenvalues, while Fig.~\ref{fig:2}~(b) shows
ground state expectation value of $X$, $Y$ and $Z$ operators.
Despite device noises in
\textbf{\textit{ibmq}}\underline{ }\textbf{\textit{lima}},
measured observables match well with exact values (dashed lines).
Moreover, computed ground state fidelity
$\mathcal{F}_\alpha=|\braket{\Phi_\alpha|\Psi_\alpha}|^2/\braket{\Psi_\alpha|\Psi_\alpha}$
(Fig.~\ref{fig:2}~(c)) is almost $1$, which
demonstrates accurate projection to the desired state by QZMC.

Figure~\ref{fig:3} presents computational results for
two-qubit systems: H$_2$ and the Hubbard dimer.
We determined the energy eigenvalues
of H$_2$ within an error of $0.02\,$Ha using
\textbf{\textit{ibm}}\underline{ }\textbf{\textit{lagos}}.
Energy eigenvalues for the Hubbard dimer
are 
calculated within an error of $0.06\,t$
on \textbf{\textit{ibm}}\underline{ }\textbf{\textit{perth}}, where $t$ is electron hopping between two hubbard atoms. 
And we compute the electronic spectral function
$A(\omega)$~\cite{Negele} of the Hubbard dimer
with the NISQ device.
Figure~\ref{fig:3}~(e)-(f) displays $A(\omega)$
at $k=0$ and $k=\pi$, showing good agreements between exact values and measured values.

The additional computations for these one- and two-qubit systems,
specifically the parameter dependence of QZMC for the one-qubit system
and the ground state energy calculation of the Hubbard dimer with
Trotterized time evolution, are provided in Sec.~\uppercase{iii}
of the Supplementary Information.

Figure~\ref{fig:4} presents the computational results for the XXZ model with $4$ to
$12$ qubits. The energy eigenvalues are well reproduced, even for $12$ qubits,
despite severe degradation of $\braket{\Psi|\Psi}$ due to device noise
and trotterization errors. Specifically, we obtained ground state energy
errors of $0.015$ for $4$ qubits, $0.0275$ for $6$ qubits, $0.016$ for $8$ qubits,
$0.041$ for $10$ qubits, and $0.051$ for $12$ qubits on
\textbf{\textit{ibm}}\underline{ }\textbf{\textit{torino}}.
These values are significantly lower than the errors in
$\braket{\Psi|\Psi}$ (represented by the differences between the squares and crosses)
shown in the right panel of the figure.
Thus, we conclude that our method provides
reasonable results even in the presence of both device noise and trotterization errors.
All calculations were performed with dynamical decoupling (DD)~\cite{Ezzell2023} and
readout error mitigation~\cite{VanDenBerg2022}, without employing
advanced techniques such as zero-noise extrapolation (ZNE)~\cite{Temme2017,Li2017,Giurgica-Tiron2020} or
probabilistic error cancellation (PEC)~\cite{VanDenBerg2022,Zhang2020}. We anticipate
that larger-scale simulations will become feasible soon with these methods
or with advancements in hardware.

\begin{figure*} 
\centering
\centerline{\includegraphics[width=17.2cm]{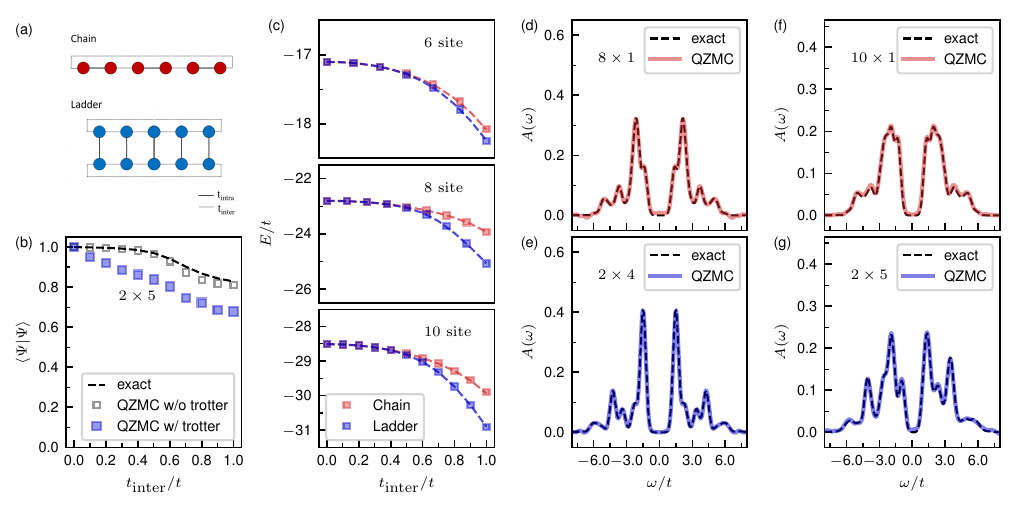}}
\caption {\textbf{The Hubbard model in various sizes}.
(a) shows two geometries we considered.
Here, colored circles denote sites,
solid lines indicate intra-dimer hopping $t_{\textrm{intra}}$,
and dotted lines represent inter-dimer hopping $t_{\textrm{inter}}$.
(b) displays $\braket{\Psi|\Psi}$ for the $2\times5$
Hubbard model as a function of $t_{\textrm{inter}}$,while
(c) presents ground state energy eigenvalues
computed from QZMC.
In each subplot of (c), red squares denote energies for $6\times1$, 
$8\times1$, and $10\times1$ models with QZMC, with red dotted lines indicating
corresponding exact values. Blue squares and lines represent the same
values for $2\times3$, $2\times4$, and $2\times5$ cases.
(d)-(g) depict the local spectral function for the Hubbard models.}
\label{fig:5}
\end{figure*}

Next, we demonstrate our method for a large system
by applying QZMC 
on the Hubbard model at the half-filling in various sizes
with noiseless qsim-cirq~\cite{qsim-cirq} quantum computer simulator. 
As $H_0$, we choose dimer array, featuring easily implementable
non-degenerate ground state.
We gradually increased the inter-dimer hopping
$t_{\textrm{inter}}$ from $0$ to the desired value $t$ as $\alpha$
increased. We explored two geometries,
chains and ladders, with periodic boundary conditions,
as illustrated in Figure~\ref{fig:5}~(a).
For each geometry, we computed systems with $6$, $8$, and $10$ sites when $U/t=5$.
For QZMC, we used $\beta=3$, with $N_\alpha$ equal to
the number of sites and $N_\nu$ increases as
$\|H'\|^2$ increases.
For the time evolution,
we used the first order Trotterization~\cite{Lloyd1996,Childs2021,Trotter1959}
, adjusting the Trotter steps as system changes.
More specifically, we used a maximum of $528$ Trotter steps for the $6 \times 1$ system and up to $1960$ steps for the $2 \times 5$ Hubbard model.

Fig.~\ref{fig:5}~(c) shows that QZMC accurately reproduces
the exact ground state energy across various configurations,
from $6$ to $10$ sites, in both chain and ladder geometries.
And our method also accurately computes
local spectral function for Hubbard models
as shown in in Fig.~\ref{fig:5}~(d)-(g),
which reproduces the exact positions and widths
of every peak in the spectral functions.
Further data not included in Fig.\ref{fig:5}(c), such as $\braket{\Psi|\Psi}$
for all geometries and spectral functions for the 6-site Hubbard models,
can be found in Sec.~\uppercase{v} of the Supplementary Information.

Finally, we computed Hubbard chains under open boundary conditions to compare our method with other methods for ground state energy estimation.
We compare our method with two state-of-the-art approaches: the Heisenberg-limited method developed by Lin and Tong~\cite{LinLin2022}, and the quantum complex exponential least squares (QCELS) method developed by Ding and Lin~\cite{Ding2023}.
We considered three cases: $4 \times 1, U = 4$; $4 \times 1, U = 10$; and $8 \times 1, U = 10$. The initial state $\ket{\tilde{\Phi}_0}$ was chosen such that $|\braket{\Phi|\tilde{\Phi}_0}|^2 = 0.4$, matching the conditions in the references~\cite{LinLin2022, Ding2023}. Both methods were implemented as described in the respective references.

\begin{figure*} 
\centering
\centerline{\includegraphics[width=17.2cm]{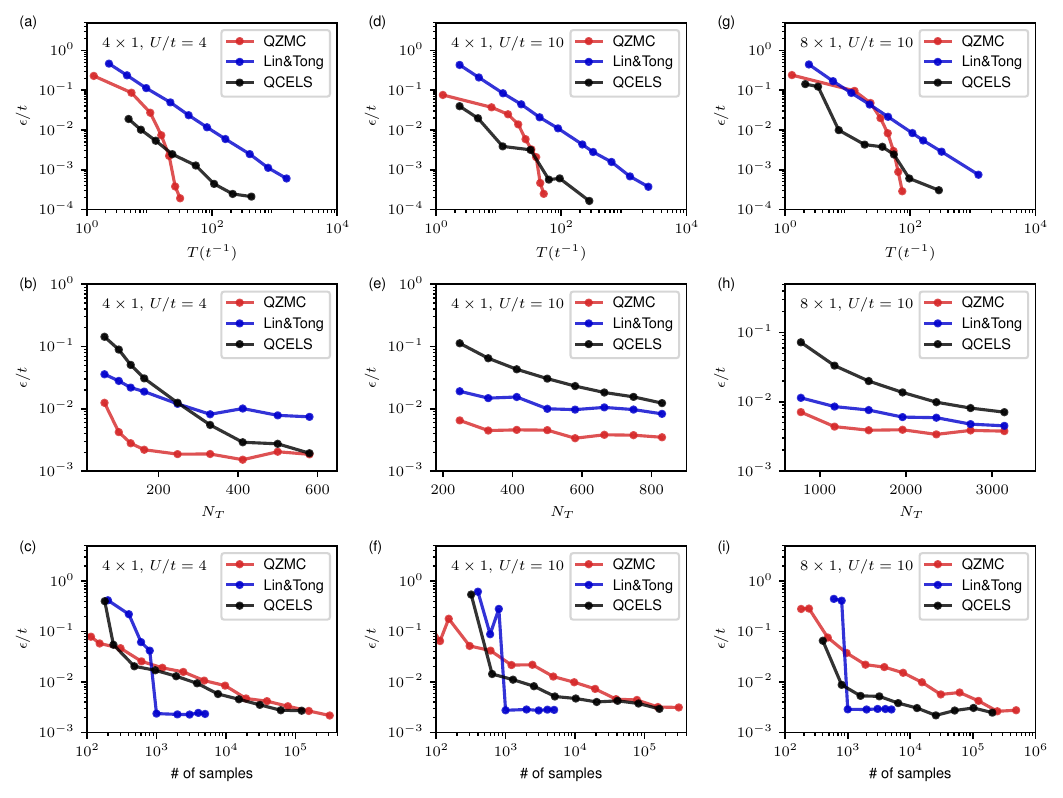}}
\caption {\textbf{Hubbard chains with various methods}.
Ground energy estimation errors are shown for: (a)-(c) 
$U/t=4$, $4$ sites; (d)-(f) $U/t=10$, $4$ sites; and (g)-(i) $U/t=10$, $8$ sites.
The figures plot the energy estimation error $\epsilon$
as a function of (a, d, g) the maximum time evolution length, (b, e, h)
the total number of Trotter steps, and (c, f, i) the total number of samples.
In all panels, blue points represent results from the method of Lin
and Tong~\cite{LinLin2022}, black points represent results
from QCELS~\cite{Ding2023}, and red points represent results from QZMC.}
\label{fig:6}
\end{figure*}

The top panels of Figure~\ref{fig:6} compares the energy estimation error $\epsilon$ as a function of the maximum time evolution length $T$.
In most of cases, QZMC requires a shorter $T$ than Lin and Tong's method and is comparable to QCELS for a precision range of $10^{-4}$ to $10^{-2}$.

The middle panels shows $\epsilon$ as a function of the total Trotterization steps $N_T$, which is directly proportional to the circuit depth. In these and the bottom panels, the maximum time evolution length $T$  
for each method was set to achieve a similar accuracy of about $0.003$ for the exact time evolution. QZMC demonstrates higher precision with fewer Trotterization steps. For example, in the $4 \times 1, U/t = 10$ case with $N_T=412$, the error for QZMC is $0.0046$, compared to $0.043$ for QCELS and $0.015$ for Lin and Tong's method.

The bottom panels plots the total number of samples required for each method. Lin and Tong's method converges quickly, while QCELS and QZMC converge more slowly, with QZMC requiring the most samples, eventually reaching approximately $10^5$. 

In conclusion, overall our method achieves higher precision with shorter circuit depth compared to other state-of-the-art methods, at the cost of requiring more samples.
Therefore, QZMC is particularly useful when quantum circuit depth is a limiting factor, but the number of accessible samples is not severely constrained.

In addition to the methods discussed above, our approach
can also be compared to adiabatic state preparation (ASP),
as both methods follow an adiabatic path. However, QZMC offers
two notable advantages over ASP. First, QZMC is resilient to
errors such as Trotter errors and device noise, making it
more practical in scenarios where such errors are significant.
Second, as highlighted in \nameref{sec:QZMC} section, QZMC does not
require the initial state $\ket{\Phi_0}$ to be exact,
whereas ASP must begin with an exact $\ket{\Phi_0}$.
This distinction is important because preparing an arbitrary
state on a quantum computer can be exponentially hard~\cite{Shende2006},
and the flexibility
to start with an approximate initial state enhances
the practicality of QZMC. A comparison of ASP and QZMC under
the influence of Trotter errors is
presented in Sec.~\uppercase{v} of the Supplementary Information.

\subsection*{Noise resilience of QZMC}
\namerefset{Noise resilience of QZMC}
\label{sec:noise_resilience}
\begin{figure} 
\centering
\centerline{\includegraphics[width=8.6cm]{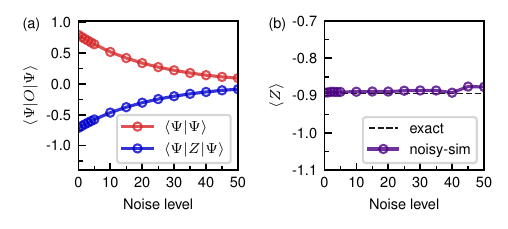}}
\caption {\textbf{Device Noise resilience of QZMC}.
$\braket{\Psi|\Psi}$, $\braket{\Psi|Z|\Psi}$, and $\braket{Z}$
of the one-qubit system considered in the Fig.~\ref{fig:2}
are drawn as a function of the noise level.
The calculations are conducted with the qiskit noisy simulator using
the noise model of \textbf{\textit{ibmq}}\underline{ }\textbf{\textit{lima}}.
In this figure, we used $N_\alpha=10$, $\beta=5$ and $N_\nu=400$.}
\label{fig:7}
\end{figure}

\begin{figure} 
\centering
\centerline{\includegraphics[width=8.6cm]{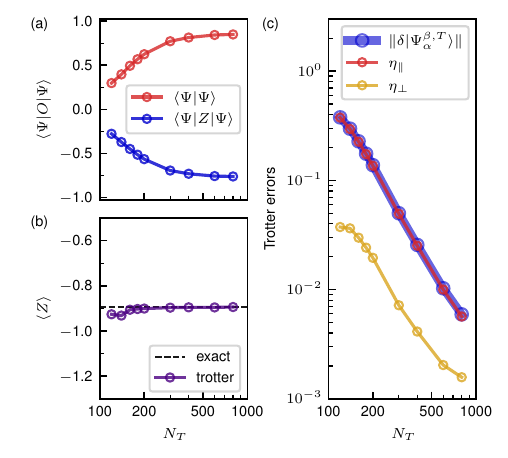}}
\caption {\textbf{Trotter Error resilience of QZMC}.
(a)-(b) show $\braket{\Psi|\Psi}$, $\braket{\Psi|Z|\Psi}$, and $\braket{Z}$ as functions of the total number of Trotter steps, for the one-qubit system considered in Fig.~\ref{fig:2}.  
In (a)-(b), $N_\nu = 400$ was used.  
(c) displays the total Trotter error $|\ket{\delta\Psi_\alpha^{\beta,T}}|$, the parallel component $\eta_\parallel$, and the orthogonal component $\eta_\perp$,
plotted against the total number of Trotter steps.
For (c), $N_\nu = 4000$ was used to reduce the statistical error.
In all panels, we set $N_\alpha = 10$ and $\beta = 5$.}
\label{fig:8}
\end{figure}

Interestingly, our calculational results for observables
accurately reproduce exact values even with the device noises (Fig.~\ref{fig:2}
and Fig.~\ref{fig:3}) and the Trotter errors (Fig.~\ref{fig:5}).
The effect of these noises induces significant
deviations of calculated $\braket{\Psi|\Psi}$ (Fig.~\ref{fig:2}~(d), Fig.~\ref{fig:3}~(c),
and Fig.~\ref{fig:5}~(b)) from exact values.
However, the observable expectation values, which is computed by using the
ratio of $\braket{\Psi|O|\Psi}$ and $\braket{\Psi|\Psi}$ (Eq.~(\ref{eq:obs}))
is robust against device noises and Trotter errors.
To understand this, we tested the dependence of the calculated observables
on the device noise magnitude using the qiskit~\cite{qiskit} aer simulator.
We considered $\braket{\Psi|\Psi}$, $\braket{\Psi|Z|\Psi}$, and $\braket{Z}$
of the ground state of the one-qubit system.
Figure.~\ref{fig:7} shows calculational results.
As the noise level increases, $\braket{\Psi|\Psi}$ decreases
and the absolute value of $\braket{\Psi|Z|\Psi}$
also decreases (Fig.~\ref{fig:7}~(a)).
Surprisingly, these noise-induced errors cancel each other through
the ratio of $\braket{\Psi|\Psi}$
and $\braket{\Psi|Z|\Psi}$, so that $\braket{Z}=\braket{\Psi|Z|\Psi}/\braket{\Psi|\Psi}$
(Fig.~\ref{fig:7}~(b)) remains robust against noise. 
Since quantum circuits for computing the
numerator and denominator are nearly identical, division
cancels out common noise effects,
making the expectation value resilient.
The same argument can be applied to Trotterization (thus,
the method is resilient to Trotter error too).
Because we use same Trotterization rule for both the
numerator and the denominator, common Trotterization
errors are canceled out by division.
This have been demonstrated numerically in 
Fig.~\ref{fig:8}~(a)-(b). In this figure,
we computed same quantities considered in Fig.~\ref{fig:7}
using trotterized time evolutions varying the total trotterization
steps $N_T$. We can see that the low-trotterization steps
makes $\braket{\Psi|\Psi}$ small, but $\braket{Z}$ does not change
a lot because the magnitude of $\braket{\Psi|Z|\Psi}$ also decreased
by the trotterization.

Figure~\ref{fig:7} and \ref{fig:8}~(a)-(b)
demonstrate that error cancellation through division occurs
in practice for both device noise and Trotter errors.
However, since these errors arise from fundamentally
different sources, the mechanisms behind their cancellation
differ. In followings, we provide a detailed
analysis of how error cancellation occurs for each type of error
and additional notes.


First, we discuss the mechanism for the device noise resilience.
In our method, we measure consecutive time evolution
using a single ancilla qubit (See Sec.~\uppercase{ii}\,D\,2
of the Supplementary Information for quantum circuits).
With this in mind, let's examine the following simple example.
Consider a qubit with the density matrix $\rho$.
Then, exact outcome of a $Z$ measurement on this qubit 
is given by $\operatorname{Tr}(\rho Z)$.
The effect of noise the qubit can be described as 
$\mathcal{E}(\rho)$~\cite{NielsenAndChuang2010}.
With this noise, the outcome of the $Z$ measurement becomes
$\operatorname{Tr}(\mathcal{E}(\rho)Z)$.
Consider the depolarizing channel as a specific type of noise,
which alters the state $\rho$ to $\mathcal{E}(\rho) = p I/2 + (1-p) \rho$.
Here, $p$ represents the probability of depolarization.
With this model, $\operatorname{Tr}(\mathcal{E}(\rho)Z)$ becomes $(1-p)\operatorname{Tr}(\rho Z)$.
Now, imagine another qubit with the density matrix $\rho'$
subjected to the same noise channel. The $Z$ measurement of this qubit
yields $(1-p)\operatorname{Tr}(\rho' Z)$.
Then, the ratio of the measurement outcomes of two qubits with noise channel
is
\begin{align}
\frac{\operatorname{Tr}(\mathcal{E}(\rho')Z)}{\operatorname{Tr}(\mathcal{E}(\rho)Z)}
= 
\frac{(1-p)\operatorname{Tr}(\rho'Z)}{(1-p)\operatorname{Tr}(\rho Z)}
=
\frac{\operatorname{Tr}(\rho'Z)}{\operatorname{Tr}(\rho Z)},
\end{align}
which is same with the exact value. This demonstrates that
the effect noise can be effectively canceled out by the division.
Though we only showed the case with the depolarizing channel,
same cancellation occurs for bit and phase flip channels.
Similar discussion can also be found in the literature on
the quantum-classical hybrid Quantum Monte Carlo algorithm (QC-QMC)~\cite{Huggins2022},
which estimates the wave function overlap efficiently using shadow tomography.

To analyze the resilience of QZMC to Trotter errors, we consider the state  
\begin{align}
    \ket{\Psi_\alpha^{\beta,T}} = \ket{\Psi_\alpha^{\beta}} + \ket{\delta\Psi_\alpha^{\beta,T}}
\end{align}  
as defined in \nameref{sec:error_analysis_eta} section.  
The error term $\ket{\delta\Psi_\alpha^{\beta,T}}$ can be decomposed into two components: one parallel to $\ket{\Psi_\alpha^{\beta}}$ and the other orthogonal to it. Suppose the error consists only of the parallel component. In this case, we can express the state as  
\begin{align}
    \ket{\Psi_\alpha^{\beta,T}} = (1 - \eta_\parallel/\|\ket{\Psi_\alpha^\beta}\|) e^{i\phi_\parallel} \ket{\Psi_\alpha^\beta}.
\end{align}  
Here, $\eta_\parallel$ represents the norm of the parallel error, and $\phi_\parallel$ is the associated phase shift.  
In such a scenario, the expectation value of an observable $O$ is  
\begin{align}
    \frac{\braket{\Psi_\alpha^{\beta,T}|O|\Psi_\alpha^{\beta,T}}}{\braket{\Psi_\alpha^{\beta,T}|\Psi_\alpha^{\beta,T}}}
    &= \frac{(1-\eta_\parallel/\|\ket{\Psi_\alpha^\beta}\|)^2\braket{\Psi_\alpha^{\beta}|O|\Psi_\alpha^{\beta}}}{(1-\eta_\parallel/\|\ket{\Psi_\alpha^\beta}\|)^2\braket{\Psi_\alpha^{\beta}|\Psi_\alpha^{\beta}}} \nonumber \\
    &= \frac{\braket{\Psi_\alpha^{\beta}|O|\Psi_\alpha^{\beta}}}{\braket{\Psi_\alpha^{\beta}|\Psi_\alpha^{\beta}}}.
\end{align}  
Thus, the parallel component of the error cancels out through division, demonstrating that QZMC is inherently resilient to this type of Trotter error.  

In practice, however, the error also contains an orthogonal component $\eta_\perp \ket{\Psi_{\alpha,\perp}^\beta}$, resulting in  
\begin{align}
    \ket{\Psi_\alpha^{\beta,T}} = (1 - \eta_\parallel/\|\ket{\Psi_\alpha^\beta}\|) e^{i\phi_\parallel} \ket{\Psi_\alpha^\beta} + \eta_\perp \ket{\Psi_{\alpha,\perp}^\beta}.
\end{align}  
Here, $\eta_\perp$ denotes the norm of the orthogonal component, and $\ket{\Psi_{\alpha,\perp}^\beta}$ is a normalized vector orthogonal to $\ket{\Psi_{\alpha}^\beta}$. Unlike the parallel component, the orthogonal error does not cancel out through division.  
Therefore, the key to Trotter error resilience lies in the relative magnitudes of $\eta_\parallel$ and $\eta_\perp$.  
Numerical tests in Figure~\ref{fig:8}~(c) demonstrate that  
$\eta_\perp \ll \eta_\parallel$ in practice.  
This dominance of the parallel component ensures that error cancellation through division remains effective,  
making the method robust against Trotter errors.

One notable point regarding noise resilience is that,
in addition to the noise cancellation effect demonstrated in Fig.\ref{fig:7}-\ref{fig:8},
the use of the estimator in Eq.(\ref{eq:corrector}) enhances robustness against noise.
This is because it computes only energy differences, limiting the
influence of noise to the energy difference $E_\alpha - E_{\alpha-1}$.
Fig.~\ref{fig:3}~(d) shows this.
In this figure, we can see that
the energy computed by Eq.~(\ref{eq:corrector})
is more precise and stable compared to the energy computed by 
$\braket{H_\alpha}_\alpha = \braket{\Phi_\alpha| H_\alpha |\Phi_\alpha}$
using Eq.~(\ref{eq:obs}).

Another important note is that our discussion on noise resilience does not imply resilience to statistical noise. In fact, as the noise level increases, the impact of statistical error on the results is amplified, requiring a larger number of samples.

\section*{Discussion}\label{sec:conclusion}
In this work, we introduced the quantum Zeno Monte Carlo (QZMC)
for the emerging stepping stone era of quantum computing~\cite{Bluvstein2024}.
This method computes static and dynamical observables of gapped quantum systems within a polynomial
quantum time, without the need for variational parameters. Leveraging the Quantum Zeno effect, we progressively approach the unknown
eigenstate from the readily solvable Hamiltonian's eigenstate.
This aspect distinguishes our method from other methods for phase estimations,
which necessitate an initial state with significant overlap with the desired eigenstate~\cite{Kitaev1995,Abrams1999,LinLin2022,Huo2023,Ding2023,Wang2023,Ni2023}.
Preparing a state with substantial overlap with an eigenstate of an easily solvable
Hamiltonian is much simpler than preparing an initial state with non-trivial overlap
with the unknown eigenstate, making our algorithm highly practical compared to other methods.
Next characteristic of the algorithm is its computation of eigenstate properties by
dividing the properties of the unnormalized eigenstate by its norm squared (Eq.~(\ref{eq:obs})).
We demonstrated that this approach effectively cancels out noise effects
in the denominator and the numerator,
rendering the method resilient to device noise as well as Trotter error.
This resilience arises from the similar noise levels experienced by
both the denominator and the numerator of observable expectation value, leading us to conclude
that our approach is well-suited for homogeneous parallel quantum computing.

\section*{Methods}\label{sec:methods}
\subsection*{NISQ simulation}
Here, we provides the details
of the NISQ simulations in Figs.~\ref{fig:2}-\ref{fig:4}.
Throughout the simulations, we used $N_s = 4000$ shots for one- and two-qubit systems, and $N_s = 2048$ shots for the XXZ model.  
Since any 1- or 2-qubit unitary operation can be represented with a small
number of gates~\cite{Shende2004},
the consecutive time evolutions encountered in QZMC can be implemented within
a shallow circuits with a few parameters.
For the 1-qubit system, the parameters $\theta_1,\theta_2,\theta_3,\theta_4$ for the unitary matrix $U$ are obtained from~\cite{Shende2004}:  
\begin{align}
    U = e^{i\theta_4} \begin{bmatrix}
    \cos(\theta_1/2) & -\sin(\theta_1/2)e^{i\theta_3} \\
    \sin(\theta_1/2)e^{i\theta_2} & \cos(\theta_1/2)e^{i(\theta_2+\theta_3)}
    \end{bmatrix}.
\end{align}  
For the 2-qubit system, we applied the two-qubit Weyl decomposition~\cite{Cross2019}, as implemented in Qiskit.  

For the XXZ model, we set $\beta = \sqrt{2}$ and combined $(P_1^{\beta})^2$ in Eq.~(\ref{eq:corrector}) into a single integral. For Trotterization, we employed second-order Trotterization based on the efficient implementation of Trotterized quantum circuits~\cite{Chowdhury2024}, using two Trotter steps.  
We begin with XXZ dimers, described by the Hamiltonian  
\begin{align}
    H_0 = -J \sum_{i;\textrm{odd}} \left(S_i^x S_{i+1}^x + S_i^y S_{i+1}^y + \Delta S_i^z S_{i+1}^z\right).
\end{align}  
For systems with up to 8 qubits, we used the first-order perturbation energy as a predictor for the energy. For the 10-qubit system, we employed $E_2 + E_8$ as the predictor, where $E_2$ is the energy of a single XXZ dimer, and $E_8$ is the energy of an 8-site XXZ model computed using \textbf{\textit{ibm}}\underline{ }\textbf{\textit{torino}}. Subsequently, using the computed $E_{10}$, we used $E_2 + E_{10}$ as the predictor for the 12-site XXZ model.  
We used an initialization circuit that prepares the vacuum state $\ket{0^n}$
when the ancilla qubit is in $\ket{0}$, and the ground state of $H_0$
when the ancilla qubit is in $\ket{1}$.
The specific initialization circuit for the 10-site XXZ model is provided in the Supplementary Information.
The number of gates used in this simulation, in terms of the basis gates of \textbf{\textit{ibm}}\underline{ }\textbf{\textit{torino}}, is 237 for 4 sites, 384 for 6 sites, 534 for 8 sites, 696 for 10 sites, and 857 for 12 sites.
\subsection*{Noiseless simulation}
Here, we discuss more detailed information
about noiseless simulations (Figures~\ref{fig:5}-\ref{fig:6}).
In these calculations, we consider the Hubbard model
which is described by the Hamiltonian
\begin{equation}
H = -\sum_{<ij>\sigma} t_{ij} c^{\dagger}_{i\sigma} c^{ }_{j\sigma} 
   - \sum_{i} \mu (n_{i\uparrow} + n_{i\downarrow})
   +  \sum_{i} U n_{i\uparrow} n_{i\downarrow},
\end{equation}
with the chemical potential $\mu=U/2$, corresponding to the half-filling.
The first two terms represent the kinetic energy and are denoted as $H_t$,
while the last term represents electron-electron interaction and is referred to as $H_U$.
The ground state of the Hubbard dimer can be expressed as
\begin{align}
    \ket{\Phi_{0,\textrm{dimer}}} = &\cos(\theta_d/2)\ket{0011} + \sin(\theta_d/2) \ket{0110} \nonumber \\
    &- \sin(\theta_d/2) \ket{1001} + \cos(\theta_d/2) \ket{1100}. \label{eq:ground_dimer}
\end{align}
Here, the angle $\theta_d$ is given by
\begin{align}
    \theta_d = -2\arctan\left(\frac{1}{2t}\left(\frac{U}{2}
               + \sqrt{\frac{U^2}{4} + 4 t^2} \right)\right). \label{eq:thetad}
\end{align}
The ground state of $H_0$,
composed of a collection of dimers, is formed by the direct product of Eq.~(\ref{eq:ground_dimer}) for each dimer.
The following describes the details specific to the calculations in Fig.~\ref{fig:5},
performed using the \emph{cirq} quantum computer simulator.
In the simulations,
we used $N_\nu$ and Trotter steps ($N_T$) that varied with the system size, while fixing the number of shots at $N_s = 10{,}000$.
Based on Eq.~(\ref{eq:minimumsampling}), $N_\nu$ was set proportional to $\|H'\|^2$, where  
\begin{align}
    \|H'\| = t \times (\textrm{number of sites}) \quad \text{(for a chain)},
\end{align}
and  
\begin{align}
    \|H'\| = \frac{4t}{\pi} \times (\textrm{number of sites}) \quad \text{(for a ladder)}.
\end{align}  
The proportionality constant was determined by testing the $6 \times 1$ system numerically.
The first-order Trotterized time evolution $U_1(\tau)$ for the Hubbard model with $n_T$ Trotter steps introduces a Trotter error~\cite{Childs2021} given by  
\begin{align}
    \|e^{-iH\tau} - U_1(\tau)\| \leq \frac{\tau^2}{2n_T} \|[H_t, H_U]\|,
\end{align}  
where  
\begin{align}
    \|[H_t, H_U]\| \leq \sum_{\langle ij \rangle\sigma} t_{ij} U \|[c^{\dagger}_{i\sigma} c^{}_{j\sigma}, \sum_{i} n_{i\uparrow} n_{i\downarrow}]\|.
\end{align}  
Since all orbital indices are equivalent, $\|[c^{\dagger}_{i\sigma} c^{}_{j\sigma}, \sum_{i} n_{i\uparrow} n_{i\downarrow}]\|$ remains constant for any $i$ and $j$. Consequently,  
\begin{align}
    \|[H_t, H_U]\| \leq C U (t_{\textrm{intra}}N_{\textrm{intra}} + t_{\textrm{inter}}N_{\textrm{inter}}),
\end{align}  
where $N_{\textrm{intra}}$ denotes the number of intra-dimer hoppings and $N_{\textrm{inter}}$ represents the number of inter-dimer hoppings,
and $C$ is a proportionality constant.

Based on this, $N_{T,\alpha}$ was determined as  
\begin{align}
    N_{T,\alpha} = \operatorname{int}\left[75 \times \frac{\left(t_{\textrm{intra}}N_{\textrm{intra}} + t_{\textrm{inter},\alpha}N_{\textrm{inter}}\right)}{8}\right],
\end{align}  
with a minimum value of $20$. Specific values of $N_\nu$ and total Trotter steps $N_T=2 \sum_{\alpha=1}^{N_\alpha} N_{T,\alpha}$ for each model are summarized
in Supplementary information.

Next, we provide detailed information on
the comparative study for the Hubbard models in Fig.~\ref{fig:6}. In this case, we considered open boundary conditions, and the initial state is prepared
from direct product of Eq.~(\ref{eq:ground_dimer}), with $\theta_d$ adjusted to achieve $ |\braket{\Phi|\tilde{\Phi}_0}|^2 = 0.4 $.  

For all data in Fig.~\ref{fig:6}, each calculation is repeated 30 times, and the absolute values of energy errors were averaged over repetitions. To measure the maximum time length $T$, we used the 99th percentile of the distribution of time evolution lengths, as all three methods are stochastic. This means that 99\% of the time evolution lengths are smaller than $T$.  

The computational parameters is set according to the references for the compared methods. For Lin and Tong's method~\cite{LinLin2022}, we set the parameter $\delta = 4/d$ as in the reference and varied $d$, which determines the time length. We used 1800 samples, consistent with the original paper. For QCELS~\cite{Ding2023}, we followed the relative gap $D$ estimation and parameter settings in the original article, using $d = \lfloor{15/D}\rfloor$ and $N = 5$. The sample number for each $n\tau_j$ was set to 2048, higher than the values used in the original paper.  

For QZMC, we used $N_\nu = 16,384$ for calculations with $\epsilon \geq 0.001$ and $N_\nu = 1,638,400$ for calculations with $\epsilon < 0.001$.
For precise calculation, after obtaining the energy difference using Eq.~(\ref{eq:corrector}), we recomputed it with the obtained $E_\alpha$ value at each $\alpha$.

For the middle and bottom panels of Fig.~\ref{fig:6},  
we noted that the maximum time evolution length $T$  
is set for each method to achieve a precision $\epsilon$  
of about $0.003$ under exact time evolution.  
In practice, the following parameters were used in our calculations.  

For the 4-site Hubbard model with $U/t = 4$, we used $d = 4000$ for Lin and Tong's method, resulting in $T = 398.56$ and $\epsilon = 2.46 \times 10^{-3}$.  
For QCELS, we used $J = 5$ and $\tau_J = 40$, yielding $T = 23.09$ and $\epsilon = 2.45 \times 10^{-3}$.  
In QZMC, we used $\beta = 1.6$, which gave $T = 20.53$ and $\epsilon = 2.23 \times 10^{-3}$.  

For the 4-site Hubbard model with $U/t = 10$, we used $d = 6000$ for Lin and Tong's method, leading to $T = 323.26$ and $\epsilon = 2.79 \times 10^{-3}$.  
In QCELS, we used $J = 7$ and $\tau_J = 108$, resulting in $T = 32.32$ and $\epsilon = 3.16 \times 10^{-3}$.  
For QZMC, we used $\beta = 2.6$, yielding $T = 33.36$ and $\epsilon = 3.24 \times 10^{-3}$.  

For the 8-site Hubbard model with $U/t = 10$, we used $d = 12000$ for Lin and Tong's method, producing $T = 316.13$ and $\epsilon = 2.84 \times 10^{-3}$.  
In QCELS, we used $J = 9$ and $\tau_J = 372$, resulting in $T = 54.65$ and $\epsilon = 2.42 \times 10^{-3}$.  
For QZMC, we used $\beta = 4.2$, giving $T = 53.88$ and $\epsilon = 2.96 \times 10^{-3}$.  

In the Trotterization tests, first-order Trotterization was employed for all methods.
In QZMC, the Trotter steps $N_{T,\alpha}$ for each $\alpha$ were determined as  
\begin{align}
    N_{T,\alpha} \propto (t_{\textrm{intra}}N_{\textrm{intra}} + t_{\textrm{inter}}N_{\textrm{inter}}),
\end{align}  
and the total Trotter steps $N_T$ were computed as $2 \sum_{\alpha} N_{T,\alpha}$.  
For calculations in Fig.~\ref{fig:6}, we used a shot number $N_s = 2048$.

\section*{Data availability}\label{sec:data_availability}
The data generated and/or analyzed during
this study are available from the corresponding author
upon reasonable request.

\section*{Code availability}\label{sec:code_availability}
The code developed during this study is available
from the corresponding author upon reasonable request.

\section*{Acknowledgment}\label{sec:acknowledgment}
We thank Hyukjoon Kwon, Joonsuk Huh, and Lin Lin for their insightful comments
and discussions. 
This research was supported by Quantum Simulator Development Project for Materials Innovation through
the National Research Foundation of Korea (NRF) funded
by the Korean government (Ministry of Science and ICT(MSIT))(No. NRF-2023M3K5A1094813).
For one and two qubit simulations,
we acknowledge the use of IBM Quantum services for this work and to
advanced services provided by the IBM Quantum Researchers
Program. The views expressed are those of the authors, and do not
reflect the official policy or position of IBM or the IBM Quantum
team.
For larger system calculation, we used resources of
the Center for Advanced Computation at Korea
Institute for Advanced Study
and
the National Energy Research
Scientific Computing Center (NERSC), a U.S. Department of Energy
Office of Science User Facility operated under
Contract No. DE-AC02-05CH11231.
SC was supported by a KIAS Individual Grant (CG090601)
at Korea Institute for Advanced Study. 
M.H. is supported by a KIAS Individual Grant (No.
CG091301) at Korea Institute for Advanced Study.

\section*{Author contributions}

M.H. conceived the original idea. M.H. and S.C. developed the idea into algorithms. M.H implemented and performed classical as well as quantum computer calculation. M.H. established the analytical proof of the computational complexity. All authors contributed to the writing of the manuscript.

\section*{Competing interests}
The authors declare no competing interests.

%

\end{document}



\title{Supplementary information for ``Quantum Zeno Monte Carlo for computing observables"}

\author{Mancheon Han}
\email{mchan@kias.re.kr}
\affiliation{School of Computational Sciences, Korea Institute for Advanced Study (KIAS), Seoul, 02455, Korea}
\author{Hyowon Park}
\email{hyowon@uic.edu}
\affiliation{Materials Science Division, Argonne National Laboratory, Argonne, IL, 60439, USA}
\affiliation{Department of Physics, University of Illinois at Chicago, Chicago, IL, 60607, USA}
\author{Sangkook Choi}
\email{sangkookchoi@kias.re.kr}
\affiliation{School of Computational Sciences, Korea Institute for Advanced Study (KIAS), Seoul, 02455, Korea}
\date{\today}

\maketitle

\section{Equations for Quantum Zeno Monte Carlo}\label{appendix:qzmc_detail}
This section provides more details of \nameref{sec:QZMC}.

\subsection{Monte Carlo Calculation of the Energy Difference}

Following the same steps used to derive Eq.~(\ref{eq:obsmc}), we obtain the Monte Carlo formula for the numerator in Eq.~(\ref{eq:corrector}):  
\begin{align}
\braket{\Psi^\beta_\alpha\ket{\Phi_\alpha}{\bra{\Phi_\alpha}}(H_\alpha-H_{\alpha-1})|\Psi^\beta_{\alpha-1}}
= \nonumber \\
\frac{1}{N_\nu}\sum_{\mathbf{t}_\nu} \langle\Phi_0| e^{-i K_1 t_{\nu,2\alpha}}
e^{-i K_2 t_{\nu,2\alpha-1}} \cdots 
e^{-i K_\alpha t_{\nu,\alpha+1}} \nonumber \\ 
e^{-i K_\alpha t_{\nu,\alpha}}
(H_\alpha-H_{\alpha-1}) 
e^{-i K_{\alpha-1} t_{\nu,\alpha-1}} \nonumber \\ 
e^{-i K_{\alpha-2} t_{\nu,\alpha-2}} \cdots 
e^{-i K_{1} t_{\nu,1}}|{\Phi_0}\rangle. \label{eq:correctormc}
\end{align}  
The measurement of the integrands on a quantum computer can be achieved using the Pauli string expansion of the operator $(H_\alpha-H_{\alpha-1})$~\cite{Tilly2022}.


\subsection{Green's function}
We derive the Quantum Zeno Monte Carlo (QZMC) method for computing the electronic Green's function. The computation of other Green's functions, such as the spin-spin correlation function, follows a similar approach. 

The retarded electronic Green's function is defined as  
\begin{align}
G^{\mathrm{R}}_{jk}(t) = - i \theta(t) \braket{\Phi|\{c_j(t), c^\dagger_k(0)\}|\Phi},
\end{align}
where $\{\, ,\, \}$ represents the anticommutator, and $c_j$ ($c^\dagger_j$) is the annihilation (creation) operator for the $j$-th orbital. This function can be expressed in terms of the spectral function $A_{jk}(\omega)$ as  
\begin{align}
G^{\mathrm{R}}_{jk}(\omega + i0^+) = \int_{-\infty}^\infty \frac{A_{jk}(\omega')}{\omega + i0^+ - \omega'} d\omega'.
\end{align}

The spectral function $A_{jk}(\omega)$ is expanded as~\cite{Negele}  
\begin{align}
A_{jk}(\omega) = A_{jk,1}(\omega) + A_{jk,2}(\omega),
\end{align} 
where  
\begin{align}
A_{jk,1}(\omega) = \sum_{m} \braket{\Phi|c_j|m}\braket{m|c^\dagger_k|\Phi} \delta(\mathcal{E}_m - \mathcal{E}_\Phi - \omega), \nonumber \\
A_{jk,2}(\omega) = \sum_{m} \braket{\Phi|c_k^\dagger|m}\braket{m|c_j|\Phi} \delta(\mathcal{E}_m - \mathcal{E}_\Phi + \omega).
\end{align} 

Here, $\ket{m}$ and $\mathcal{E}_m$ denote the energy eigenstates and their corresponding eigenvalues. We focus on computing $A_{jk,1}(\omega)$, as $A_{jk,2}(\omega)$ can be computed similarly. The Dirac delta function $\delta(x)$ is approximated using a Gaussian function:  
\begin{align}
\delta(x) \approx \frac{1}{\sqrt{2\pi\sigma^2}} e^{-x^2/2\sigma^2}.
\end{align}
With Gaussian broadening, $A_{jk,1}(\omega)$ is computed as  
\begin{align}
A_{jk,1}(\omega) = \frac{1}{\sqrt{2\pi\sigma^2}} \braket{\Phi|c_j e^{-(H - \mathcal{E}_\Phi - \omega)^2/2\sigma^2} c^\dagger_k|\Phi}.
\end{align}

We consider the Green's function at $\alpha$. Similar to Eq.~(\ref{eq:obs}), $A_{jk,1}(\omega)$ can be expressed as  
\begin{align}
A_{jk,1}(\omega) = \frac{1}{\sqrt{2\pi\sigma^2}} \frac{\tilde{A}_{jk,1}(\omega)}{\braket{\Psi_\alpha|\Psi_\alpha}},
\end{align}
where  
\begin{align}
\tilde{A}_{jk,1}(\omega) = \braket{\Psi_\alpha|c_j e^{-(H_\alpha - E_\alpha - \omega)^2/2\sigma^2} c^\dagger_k|\Psi_\alpha}. 
\end{align}

Using a Fourier expansion of $e^{-(H_\alpha - E_\alpha - \omega)^2}$, $\tilde{A}_{jk,1}(\omega)$ can be computed via Monte Carlo sampling, similar to Eq.~(\ref{eq:obsmc}):  
\begin{align}
\tilde{A}_{jk,1}(\omega) = \frac{1}{N_\nu} \sum_\nu g_{\alpha}(c_j, c^\dagger_k, \mathbf{t}_\nu) e^{i\phi_\alpha(\mathbf{t}_\nu, \omega)},\label{eq:greenmc}
\end{align}
where  
\begin{align}
g_{\alpha}(O_1, O_2, \mathbf{t}) = \langle \Phi_0| e^{-i H_1 t_{2\alpha+1}} 
e^{-i H_2 t_{2\alpha}} \cdots 
\nonumber \\
e^{-i H_\alpha t_{\alpha+2}}
O_1 e^{-i H_\alpha t_{\alpha+1}} O_2 e^{-i H_\alpha t_\alpha} 
\nonumber \\
e^{-i H_{\alpha-1} t_{\alpha-1}} 
\cdots e^{-i H_1 t_1}|\Phi_0
\rangle, 
\end{align}
and $  \phi_\alpha(\mathbf{t}, \omega) = \sum_{\alpha'=1}^\alpha E_{\alpha'} (t_{\alpha'} + t_{2\alpha+2-\alpha'}) + (E_\alpha + \omega) t_{\alpha+1}$.
Here, $t_{\nu,\alpha+1} \sim \mathcal{N}(0, 1/\sigma)$,
and $t_{\nu,\alpha'} \sim \mathcal{N}(0, \beta)$ for all $\alpha' \neq \alpha+1$. Since $g_{\alpha}(O_1, O_2, \mathbf{t})$ is independent of $\omega$, $A_{jk}(\omega)$ can be computed for any $\omega$ by storing $g_{\alpha}(c_j, c^\dagger_k, \mathbf{t}_\nu)$.

\section{Supplement to the error analysis and cost estimation\label{appendix:cost}}
This section provides additional details on Error analysis
and Cost estimation in the main text.
As noted in the main text, we consider the
Hamiltonian $H(\lambda) = H_0 + \lambda H'$,
starting with $\ket{\Phi_0}$, an eigenstate
of $H_0 = H(\lambda=0)$ with eigenvalue $E_0$.
For simplicity, we assume that $\ket{\Phi(\lambda)}$,
which is adiabatically connected to $\ket{\Phi_0}$,
is non-degenerate and has an eigenvalue $E(\lambda)$
satisfying $|E(\lambda) - \mathcal{E}_m(\lambda)| \geq \Delta_g$
for all other eigenenergies $\mathcal{E}_m(\lambda)$ of $H(\lambda)$.
We then define $H_\alpha = H(\lambda_\alpha)$ for $\alpha = 1, 2, \dots, N_\alpha$,
with $\lambda_\alpha = \alpha / N_\alpha$, as described in the main text.
For simplicity, we denote $E_\alpha = E(\lambda_\alpha)$
and $\ket{\Phi_\alpha} = \ket{\Phi(\lambda_\alpha)}$.
The eigenvectors of $ H_\alpha$
are denoted by $\ket{m_\alpha}$,
with corresponding eigenvalues $\mathcal{E}_{m_\alpha}$.
In particular, $\mathcal{E}_{m_\alpha} = E_\alpha$
when $m_\alpha = \Phi_\alpha$.

\subsection{Error analysis of $\braket{\Psi^\beta_\alpha|\Psi^\beta_\alpha}$}
\subsubsection{The number of projections}
First, we derive Eq.~(\ref{eq:proj1}) in the main text, 
\begin{align}
|\braket{\Phi_\alpha|\Phi_{\alpha+1}}|^2 \geq 1-\|H'\|^{2}\Delta_g^{-2}N_\alpha^{-2}, \label{eq:prove_1}
\end{align}  
using the perturbation theory~\cite{LandauQM}.
Starting from $H_{\alpha+1} = H_\alpha + \Delta\lambda H'$, we expand $\ket{\Phi_{\alpha+1}}$ as  
\begin{align}
    \ket{\Phi_{\alpha+1}} = \ket{\Phi_\alpha} + (\Delta\lambda) \ket{\Phi_\alpha^{(1)}} + 
\frac{(\Delta\lambda)^2}{2} \ket{\Phi_\alpha^{(2)}},
\end{align}  
omitting higher-order terms. Similarly, with $E_{\alpha+1} = E_\alpha + \Delta\lambda E'_\alpha + (\Delta\lambda)^2/2 E''_\alpha + \cdots$, and collecting $\Delta\lambda$-order terms from $H_{\alpha+1}\ket{\Phi_{\alpha+1}} = E_{\alpha+1}\ket{\Phi_{\alpha+1}}$, we obtain  
\begin{align}
    (E_\alpha-H_\alpha) \ket{\Phi^{(1)}_\alpha} = (H'-E'_\alpha) \ket{\Phi_\alpha}.
\end{align}  
Taking the inner product with $\ket{m_\alpha}$ on both sides gives  
\begin{align}
    \braket{m_{\alpha}|\Phi_{\alpha}^{(1)}} = \frac{\braket{m_\alpha|H'|\Phi_\alpha}}{E_\alpha-\mathcal{E}_{m,\alpha}}, \quad \text{for } m_\alpha \neq \Phi_\alpha.
\end{align}  

Since $\ket{\Phi_{\alpha+1}}$ is normalized,
$\braket{\Phi_{\alpha+1}|\Phi_{\alpha+1}}$ must not include $\Delta\lambda$-dependent terms.
And to uniquely define the phase of $\ket{\Phi_{\alpha+1}}$, we set $\braket{\Phi_{\alpha+1}|\Phi_\alpha}$ to be real. Then,
\begin{align}
    \braket{\Phi^{(1)}_\alpha|\Phi_\alpha} + \braket{\Phi_\alpha|\Phi^{(1)}_\alpha} = 2 \braket{\Phi_\alpha|\Phi^{(1)}_\alpha} = 0 \nonumber, \\
    \braket{\Phi_\alpha|\Phi^{(2)}_\alpha} + \braket{\Phi^{(1)}_\alpha|\Phi^{(1)}_\alpha} = 0.
\end{align}  

This yields  
\begin{align}
    \braket{\Phi_\alpha|\Phi^{(2)}_\alpha} = -\sum_{m_\alpha\neq\Phi_\alpha}
    \frac{|\braket{m_\alpha|H'|\Phi_\alpha}|^2}{(E_\alpha-\mathcal{E}_{m,\alpha})^2}.
\end{align}  

Using this result, we have  
\begin{align}
    \braket{\Phi_{\alpha}|\Phi_{\alpha+1}} &= 1-\frac{(\Delta\lambda)^2}{2}\sum_{m_\alpha\neq\Phi_\alpha}
    \frac{|\braket{m_\alpha|H'|\Phi_\alpha}|^2}{(E_\alpha-\mathcal{E}_{m,\alpha})^2},\nonumber \\
    \braket{m_{\alpha}|\Phi_{\alpha+1}} &= \Delta\lambda \frac{\braket{m_\alpha|H'|\Phi_\alpha}}{E_\alpha-\mathcal{E}_{m,\alpha}}, \quad \text{for } m_\alpha \neq \Phi_\alpha, \label{eq:perturb}
\end{align}  
to leading order in $\Delta\lambda = \lambda_{\alpha+1} - \lambda_\alpha$. Since $|\mathcal{E}_{m,\alpha} - E_\alpha| \geq \Delta_g$,  
\begin{align}
 \sum_{m_\alpha\neq\Phi_\alpha}\frac{|\braket{m_\alpha|H'|\Phi_\alpha}|^2}{(E_\alpha-\mathcal{E}_{m,\alpha})^2}
 \leq\frac{1}{\Delta_g^2}\|H'\|^2. \label{eq:coeffbound}
\end{align}  
Here, we use the matrix norm induced by the vector 2-norm~\cite{HeathSC}, defined as $\|A\| = \sup_{\|x\|=1} \|Ax\|$. Combining Eq.~(\ref{eq:perturb}) and Eq.~(\ref{eq:coeffbound}),
we obtain Eq.~(\ref{eq:prove_1}).   

Next, we consider the ground state. Returning to perturbative analysis and collecting $(\Delta\lambda)^2$-order terms in  
$\braket{\Phi_\alpha|H_{\alpha+1}|\Phi_{\alpha+1}} = E_{\alpha+1}\braket{\Phi_\alpha|\Phi_{\alpha+1}}$, we find  
\begin{align}
    \frac{1}{2} E''_\alpha =  \braket{\Phi_\alpha |H'|\Phi_{\alpha}^{(1)}} =
    \sum_{m_\alpha\neq\Phi_\alpha}
    \frac{|\braket{m_\alpha|H'|\Phi_\alpha}|^2}{(E_\alpha-\mathcal{E}_{m,\alpha})}.
\end{align}  
For a ground state, $|E_\alpha - \mathcal{E}_{m,\alpha}| \geq \Delta_g$
gives $E_\alpha-\mathcal{E}_{m,\alpha}\leq -\Delta_g$, $(E_\alpha-\mathcal{E}_{m,\alpha})^2 \geq -\Delta_g (E_\alpha-\mathcal{E}_{m,\alpha})$.
Thus,  
\begin{align}
\braket{\Phi_{\alpha}|\Phi_{\alpha+1}} &= 1 - \frac{(\Delta\lambda)^2}{2}\sum_{m_\alpha\neq\Phi_\alpha}
\frac{|\braket{m_\alpha|H'|\Phi_\alpha}|^2}{(E_\alpha-\mathcal{E}_{m,\alpha})^2}\nonumber\\
&\geq 1 + \frac{(\Delta\lambda)^2}{4}\frac{E''(\lambda_\alpha)}{\Delta_g}.
\end{align}  

This implies  
\begin{align}
    \braket{\Psi_\alpha|\Psi_\alpha} &\geq 1 + \frac{1}{\Delta_g}\sum_{\alpha} \frac{E''(\lambda_\alpha)}{2N_\alpha^2} \nonumber \\
    &= 1 + \frac{1}{\Delta_g}\frac{(E'(1)-E'(0))}{2N_\alpha} + \mathcal{O}\left(\frac{1}{N^2_\alpha}\right).
\end{align}  
Since $|E'(\lambda)| = |\braket{\Phi(\lambda)|H'|\Phi(\lambda)}| \leq \|H'\|$, we conclude  
\begin{align}
    \braket{\Psi_\alpha|\Psi_\alpha} \geq 1-\frac{\|H'\|}{\Delta_g}\frac{1}{N_\alpha},
\end{align}  
to leading order in $1/N_\alpha$. Therefore, for the ground state, it is sufficient to have  
\begin{align}
    N_\alpha \geq \frac{\|H'\|}{\Delta_g}\frac{1}{\eta_0}, \label{eq:N_alpha_ground}
\end{align}  
to ensure $\|\ket{\Psi_\alpha}\|^2 \geq 1-\eta_0$.

\subsubsection{The projection errors}
As defined in the main text, the error in the projected state is  
\begin{align}
    \ket{\delta\Psi^\beta_\alpha} = \ket{\Psi^\beta_\alpha} - \ket{\Psi_\alpha} = (\mathcal{P}^\beta_\alpha-\mathcal{P}_\alpha)\ket{\Phi_0}.
\end{align}  
Here, we show Eq.~(\ref{eq:betabound}) in the main text,  
\begin{align}
\|\ket{\delta\Psi^\beta_\alpha}\|\leq \frac{\alpha}{N_\alpha}e^{-\beta^2\Delta_g^2/2}\frac{\|H'\|}{\Delta_g}, \label{eq:beta_norm}
\end{align}  
using an inductive argument. For convenience, we define $\delta\mathcal{P}_\alpha^\beta = \mathcal{P}_\alpha^\beta - \mathcal{P}_\alpha$, $P_\alpha^\beta = P^\beta_{H_\alpha}(E_\alpha)$, and $\delta P_\alpha^\beta = P^\beta_{\alpha}-\ket{\Phi_\alpha}\bra{\Phi_\alpha}$.
Then, up to the leading order in $1/N_\alpha$, we have  
\begin{align}
    &\|\ket{\delta P_{\alpha}^\beta |\Psi_{\alpha-1}}\|^2 = \braket{\Psi_{\alpha-1}|(\delta P_{\alpha}^\beta)^2 |\Psi_{\alpha-1}} \nonumber \\
                             &\leq (\Delta\lambda)^2 \sum_{m_\alpha\neq\Phi_\alpha}
                             e^{-\beta^2(\mathcal{E}_{m,\alpha}-E_{\alpha})^2}\frac{|\braket{\Phi_{\alpha}|H'|m_{\alpha}}|^2}{(\mathcal{E}_{m,\alpha}-E_{\alpha})^2},
\end{align}  
by applying perturbation from $H_\alpha$ to $H_{\alpha-1}$, which results in Eq.~(\ref{eq:perturb}) but with a reversed sign for $\Delta\lambda$. Because $\exp(-\beta^2 x^2)/x^2$ is a decreasing function, we obtain  
\begin{align}
\|\ket{\delta P_{\alpha}^\beta |\Psi_{\alpha-1}}\| \leq \frac{1}{N_\alpha} e^{-\beta^2\Delta_g^2/2} \frac{\|H'\|}{\Delta_g}. \label{eq:beta_lemma}
\end{align}  

For $\alpha=1$, this directly gives Eq.~(\ref{eq:beta_norm}). Now, suppose that for some $\alpha$,  
\[
\|\ket{\delta\Psi^\beta_\alpha}\| \leq \frac{\alpha}{N_\alpha}e^{-\beta^2\Delta_g^2/2}\frac{\|H'\|}{\Delta_g}.
\]  

Then, using  
\begin{align}
    \ket{\delta\Psi^\beta_{\alpha+1}} = \delta P^\beta_{\alpha+1}\ket{\Psi_\alpha} + P^\beta_{\alpha+1}\ket{\delta\Psi^\beta_\alpha},
\end{align}  
and the inequality $\|v_1 + v_2\| \leq \|v_1\| + \|v_2\|$, we have  
\begin{align}
    \|\ket{\delta\Psi^\beta_{\alpha+1}}\| &\leq \|\delta P^\beta_{\alpha+1} \ket{\Psi_\alpha}\| + \|P^\beta_{\alpha+1} \ket{\delta \Psi_\alpha}\| \nonumber \\
                                          &\leq \frac{\alpha+1}{N_\alpha} e^{-\beta^2\Delta_g^2/2} \frac{\|H'\|}{\Delta_g},
\end{align}  
where the second inequality follows from Eq.~(\ref{eq:beta_lemma}). This proves Eq.~(\ref{eq:beta_norm}).

\subsubsection{The trotterization errors}
Let us define $\ket{\psi_\alpha(\mathbf{t})} = \exp(-iH_\alpha t_\alpha)\exp(-iH_{\alpha-1}t_{\alpha-1}) \allowbreak\cdots \exp(-iH_1t_1)\ket{\Phi_0}$ and $\ket{\psi^T_\alpha(\mathbf{t})}$ as its trotterized version, such that  
$\ket{\psi^T_\alpha(\mathbf{t})} = \tilde{U}_\alpha(t_\alpha)\tilde{U}_{\alpha-1}(t_{\alpha-1}) \cdots \tilde{U}_1(t_1)\ket{\Phi_0}$,
where $\tilde{U}_\alpha(t_\alpha)$ is the Trotterized approximation of $\exp(-iH_\alpha t_\alpha)$. The difference $\ket{\delta\Psi_\alpha^{\beta,T}} = \ket{\Psi_\alpha^{\beta,T}} - \ket{\Psi_\alpha^{\beta}}$
can be expressed as  
\begin{align}
    \ket{\delta\Psi_\alpha^{\beta,T}} = \frac{1}{(2\pi\beta^2)^{\alpha/2}} \int e^{-\sum_{\alpha'=1}^{\alpha} t_{\alpha'}^2/(2\beta^2) + i\phi_\alpha(\mathbf{t})} \nonumber \\
    \times (\ket{\psi^T_\alpha(\mathbf{t})} - \ket{\psi_\alpha(\mathbf{t})}) d\mathbf{t}, \label{eq:trotter_expansion}
\end{align}  
while $\phi_\alpha = \sum_{\alpha'=1}^{\alpha} E_{\alpha'} t_{\alpha'}$.
Letting $\delta\tilde{U}_{\alpha}(t_{\alpha}) = \tilde{U}_{\alpha}(t_{\alpha}) - \exp(-iH_{\alpha} t_{\alpha})$, we have~\cite{Childs2021}:  
\begin{align}
    \|\delta\tilde{U}_{\alpha}(t_{\alpha})\| \leq C_{\alpha,p} \frac{|t_{\alpha}|^{1+p}}{N_{T,\alpha}^p},
\end{align}  
where $p$ is the Trotterization order, $N_{T,\alpha}$ is the number of Trotter steps for the $\alpha$-th evolution,
and $C_{\alpha,p}$ is
the coefficient that is proportional to the sum of the norms of the commutators~\cite{Childs2021}.
Thus,  
\begin{align}
    \|\delta\tilde{U}_\alpha(t_\alpha)\ket{\psi_{\alpha-1}(\mathbf{t})}\| \leq C_{\alpha,p} \frac{|t_\alpha|^{1+p}}{N_{T,\alpha}^p}.
\end{align}  

Using an inductive argument similar to that in the proof of Eq.~(\ref{eq:beta_norm}), we find  
\begin{align}
    \|\ket{\psi^T_\alpha(\mathbf{t})} - \ket{\psi_\alpha(\mathbf{t})}\| \leq \sum_{\alpha'=1}^\alpha C_{\alpha',p} \frac{|t_{\alpha'}|^{1+p}}{N_{T,\alpha'}^p}.
\end{align}  

Substituting this into Eq.~(\ref{eq:trotter_expansion}), we have  
\begin{align}
    \|\ket{\delta \Psi_\alpha^{\beta,T}}\| \leq & \sum_{\alpha'=1}^\alpha C_{\alpha',p} \frac{M_{1+p}(\beta)}{N_{T,\alpha'}^p}, \nonumber \\
    M_{1+p}(\beta) = & \frac{1}{\sqrt{2\pi\beta^2}} \int_{-\infty}^\infty e^{-t^2/(2\beta^2)} |t|^{1+p} dt. \label{eq:trotter_norm_0}
\end{align}  

By choosing $N_{T,\alpha}$ to be proportional to $C_{\alpha,p}^{1/p} M^{1/p}_{1+p}(\beta)$, we have  
\begin{align}
    N_{T,\alpha} = \frac{C_{\alpha,p}^{1/p} M^{1/p}_{1+p}(\beta)}{\sum_{\alpha'=1}^{N_\alpha} C_{\alpha',p}^{1/p} M^{1/p}_{1+p}(\beta)} \frac{N_T}{2},
\end{align}  
while total number of trotter steps $N_T = \sum_{\alpha=1}^{N_\alpha} N_{T,\alpha}$.
Then,
\begin{align}
    \|\ket{\delta \Psi_\alpha^{\beta,T}}\| \leq & \alpha \frac{
        \left(\sum_{\alpha'=1}^{N_\alpha} C_{\alpha',p}^{1/p} M^{1/p}_{1+p}(\beta)\right)^p}
        {(N_T/2)^p}. \label{eq:trotter_norm}
\end{align}  
By setting this bound smaller than $\eta_T/2$ at $\alpha=N_\alpha$,
we obtain Eq.~(\ref{eq:trottercondition}).

\subsection{Error analysis of $E_\alpha$}
\subsubsection{The projection errors}
The numerator in Eq.~(\ref{eq:corrector}) has an error of $\Delta\epsilon_\beta \braket{\Psi_\alpha^\beta|\Psi_\alpha^\beta}$, given by  
\begin{align}
\Delta\epsilon_\beta \braket{\Psi_\alpha^\beta|\Psi_\alpha^\beta}=
    \braket{\Psi^\beta_{\alpha-1}|(P^\beta_{\alpha})^2 \Delta\lambda 
    H' |\Psi^\beta_{\alpha-1}} \nonumber \\
    -\braket{\Psi^\beta_{\alpha-1}|
    (P^\beta_{\alpha})^2(E_{\alpha}-E_{\alpha-1})|\Psi^\beta_{\alpha-1}}. \label{eq:e_proj_error_1}
\end{align}  
Substituting $\ket{\Psi^\beta_{\alpha-1}} = \ket{\Psi_{\alpha-1}} + \ket{\delta\Psi^\beta_{\alpha-1}}$, we can expand Eq.~(\ref{eq:e_proj_error_1}) as  
\begin{align}
    \Delta\epsilon_\beta \braket{\Psi_\alpha^\beta|\Psi_\alpha^\beta} =
    \braket{\delta\Psi^\beta_{\alpha-1}|(P^\beta_{\alpha})^2 \Delta\lambda 
    H' |\Psi_{\alpha-1}} \nonumber  \\ 
    +\braket{\Psi_{\alpha-1}|(P^\beta_{\alpha})^2 \Delta\lambda 
    H' |\delta\Psi^\beta_{\alpha-1}} \nonumber \\
    -\braket{\delta\Psi^\beta_{\alpha-1}|(P^\beta_{\alpha})^2 
    (E_\alpha-E_{\alpha-1})
    |\Psi_{\alpha-1}} \nonumber \\
    -\braket{\Psi_{\alpha-1}|(P^\beta_{\alpha})^2 
    (E_\alpha-E_{\alpha-1})
    |\delta\Psi^\beta_{\alpha-1}},
\end{align}  
up the the leading order of $|\ket{\delta\Psi_{\alpha-1}^\beta}|$.
From Eq.~(\ref{eq:beta_norm}), the first and second terms are bounded above by $(\alpha-1)/N_\alpha^2 \exp(-\beta^2\Delta_g^2/2) \|H'\|^2\Delta_g^{-1}$.  
For the third and fourth terms, by the mean value theorem,
there exists a $\lambda \in (\lambda_{\alpha-1}, \lambda_\alpha)$ such that $E_\alpha-E_{\alpha-1} = \braket{\Phi(\lambda)|H'|\Phi(\lambda)} \Delta\lambda$.
Because $|\braket{\Phi(\lambda)|H'|\Phi(\lambda)}| \leq \|H'\|$,
the third and fourth terms have the same bound as the first and second terms, resulting in  
\begin{align}
    \Delta\epsilon_\beta \braket{\Psi_\alpha^\beta|\Psi_\alpha^\beta} \leq \frac{4 (\alpha-1)}{N_\alpha^2} e^{-\beta^2\Delta_g^2/2} \frac{\|H'\|^2}{\Delta_g},
\end{align}  
up to the leading order of $1/N_\alpha$.  
Since $\braket{\Psi_\alpha^\beta|\Psi_\alpha^\beta} \geq 1-\eta$ and $\epsilon_\beta = \sum_{\alpha=1}^{N_\alpha} \Delta\epsilon_\beta$,
we obtain  
\begin{align}
    \epsilon_\beta \leq 2 e^{-\beta^2\Delta_g^2/2} \frac{\|H'\|^2}{\Delta_g}\frac{1}{1-\eta}.
\end{align}

\subsubsection{The trotterization errors}
Following the same reasoning as in the previous discussion on $\beta$ and from Eq.~(\ref{eq:trotter_norm}), we derive the bound for the error term $\Delta\epsilon_T \braket{\Psi_\alpha^{\beta,T}|\Psi_\alpha^{\beta,T}}$ as  
\begin{align}
    \Delta\epsilon_T \braket{\Psi_\alpha^{\beta,T}|\Psi_\alpha^{\beta,T}} \leq
    4 \frac{\alpha-1}{N_\alpha}
        \frac{\left(\sum_{\alpha'=1}^{N_\alpha} C_{\alpha',p}^{1/p} M^{1/p}_{1+p}(\beta)\right)^p}
        {(N_T/2)^p} \|H'\|,
\end{align}  
Summing over $\alpha$, the total Trotterization error is  
\begin{align}
    \epsilon_T = \sum_{\alpha=1}^{N_\alpha} \Delta\epsilon_T.
\end{align}  
Using the bound for $\Delta\epsilon_T$, we obtain  
\begin{align}
    \epsilon_T \leq 2 N_\alpha \left(\frac{N_T}{2}\right)^{-p}
        \left(\sum_{\alpha=1}^{N_\alpha} C_{\alpha,p}^{1/p} M^{1/p}_{1+p}(\beta)\right)^p
        \frac{\|H'\|}{1-\eta}.
\end{align}


\subsubsection{The sampling errors}
Defining $g(\mathbf{t}) = (2\pi \beta^2)^{-\alpha} e^{-(t_1^2 + t_2^2 + \dots + t_{2\alpha}^2)/(2\beta^2)}$, we express  
\begin{align}
    y(\mathbf{t}) = &
    \langle \Phi_0 | e^{-iK_1 t_{2\alpha}} e^{-iK_2 t_{2\alpha-1}} \cdots e^{-iK_{\alpha} t_{\alpha}}
    (H_\alpha - H_{\alpha-1}) \nonumber \\
    & e^{-iK_{\alpha-1} t_{\alpha-1}} e^{-iK_{\alpha-2} t_{\alpha-2}} \cdots e^{-iK_1 t_1} | \Phi_0 \rangle,
\end{align}  
and  
\begin{align}
    x(\mathbf{t}) = 
    \langle \Phi_0 | e^{-iK_1 t_{2\alpha}} e^{-iK_2 t_{2\alpha-1}} \cdots e^{-iK_{\alpha} t_{\alpha}} \nonumber \\
    e^{-iK_{\alpha-1} t_{\alpha-1}} e^{-iK_{\alpha-2} t_{\alpha-2}} \cdots e^{-iK_1 t_1} | \Phi_0 \rangle.
\end{align}  
The computation of Eq.~(\ref{eq:corrector}) can then be viewed as finding the ratio of the expectation values of $y(\mathbf{t})$ and $x(\mathbf{t})$ under the probability distribution $g(\mathbf{t})$. Using $N_\nu$ samples, the averages $\bar{x}$ and $\bar{y}$ are estimators of $x(\mathbf{t})$ and $y(\mathbf{t})$, respectively, with $\bar{y}/\bar{x}$ serving as the estimator for the energy difference.  

Since each term in $y(\mathbf{t})$ is bounded by $\Delta\lambda \|H'\|$, we have
$\|\bar{y}\| \leq \Delta\lambda \|H'\|$,\, $\sigma^2_y \leq (\Delta\lambda)^2 \|H'\|^2$, and the variance of the estimator satisfies $\sigma^2_{\bar{y}} \leq (\Delta\lambda)^2\|H'\|^2/N_\nu$. Similarly, for $x(\mathbf{t})$, $\sigma^2_{\bar{x}} \leq 1/N_\nu$.  
Using the delta method~\cite{Kroese2011} and assuming independent sampling of $\mathbf{t}$ for $\bar{x}$ and $\bar{y}$, the variance of the ratio $\bar{y}/\bar{x}$ is given by  
\begin{align} 
    \sigma^2_{\bar{y}/\bar{x}} = \frac{1}{\bar{x}^2} \left( \sigma^2_{\bar{y}} + \frac{\bar{y}^2}{\bar{x}^2} \sigma^2_{\bar{x}} \right) \leq \frac{(\Delta\lambda)^2 \|H'\|^2}{N_\nu \bar{x}^2} \left( 1 + \frac{1}{\bar{x}^2} \right).
\end{align}  
Thus, the standard error $\sigma_{E_\alpha - E_{\alpha-1}}$ for the energy difference computed using Eq.~(\ref{eq:correctormc}) satisfies
\begin{align}
    \sigma_{E_\alpha - E_{\alpha-1}} \leq \Delta\lambda \frac{\|H'\|}{\sqrt{N_\nu}} \frac{1}{\braket{\Psi_\alpha^\beta | \Psi_\alpha^\beta}} \sqrt{1 + \frac{1}{\braket{\Psi_\alpha^\beta | \Psi_\alpha^\beta}^2}}. \label{eq:montecarlo_error_corrector}
\end{align}  
The cumulative error for the energy estimation is then  
\begin{align}
    \sigma_{E_\alpha} = \sigma_{E_\alpha - E_0} \leq \sum_{\alpha'=1}^\alpha \sigma_{E_{\alpha'} - E_{\alpha'-1}}.
\end{align}  
Using Eq.~(\ref{eq:montecarlo_error_corrector}) and $\braket{\Psi_\alpha^\beta|\Psi_\alpha^\beta} \geq 1-\eta$,
\begin{align}
    \sigma_{E_\alpha} \leq \frac{\alpha}{N_\alpha} \frac{\|H'\|}{\sqrt{N_\nu}} \frac{1}{1-\eta} \sqrt{1 + \frac{1}{(1-\eta)^2}}.
\end{align}  
Finally, by setting $\alpha = N_\alpha$ and $\epsilon_{mc} = \sigma_{E_\alpha}$, we obtain Eq.~(\ref{eq:energy_sampling_bound}).

\subsection{Error analysis and cost estimation for $\braket{O}$}

In this section, we analyze error of $\braket{O}$ estimated from QZMC.
The error of $\braket{O}$
is decomposed as $\epsilon_\beta + \epsilon_T + \epsilon_\textrm{mc}$,
while, $\epsilon_\beta$ arises from the finite $\beta$,
$\epsilon_T$ is due to trotterization, and $\epsilon_\textrm{mc}$
results from the finite number of samples.

\subsubsection{The projection errors}
From $\ket{\Psi^\beta_\alpha} = \ket{\Psi_\alpha} + \ket{\delta\Psi^\beta_\alpha}$, the expectation value $\braket{\Psi^\beta_\alpha|O|\Psi^\beta_\alpha}$ has an error given by  
\begin{align}
    \delta{\braket{\Psi^\beta_\alpha|O|\Psi^\beta_\alpha}} = \braket{\delta\Psi_\alpha^\beta|O|\Psi_\alpha} + \braket{\Psi_\alpha|O|\delta\Psi_\alpha^\beta} \nonumber \\
    + \braket{\delta\Psi^\beta_\alpha|O|\delta\Psi_\alpha^\beta},
\end{align}  
which is bounded by  
\begin{align}
    2\|\ket{\delta\Psi^\beta_\alpha}\|\|O\| + \|\ket{\delta\Psi^\beta_\alpha}\|^2\|O\|.
\end{align}  
Considering errors up to the first order of $\|\ket{\delta\Psi^\beta_\alpha}\|$, we simplify the bound to  
\begin{align}
    \delta{\braket{\Psi^\beta_\alpha|O|\Psi^\beta_\alpha}} \leq 2\|\ket{\delta\Psi^\beta_\alpha}\|\|O\|.
\end{align}  
The error $\epsilon_\beta$ in the expectation value $\braket{\Psi^\beta_\alpha|O|\Psi^\beta_\alpha}/\braket{\Psi^\beta_\alpha|\Psi^\beta_\alpha}$ is then estimated as  
\begin{align}
    \delta{\left(\frac{\braket{\Psi^\beta_\alpha|O|\Psi^\beta_\alpha}}{\braket{\Psi^\beta_\alpha|\Psi^\beta_\alpha}}\right)} &= \frac{\delta\braket{\Psi^\beta_\alpha|O|\Psi^\beta_\alpha}}{\braket{\Psi_\alpha|\Psi_\alpha}} + \frac{\braket{\Psi_\alpha|O|\Psi_\alpha}}{\braket{\Psi_\alpha|\Psi_\alpha}^2} \delta{\braket{\Psi^\beta_\alpha|\Psi^\beta_\alpha}}, \nonumber \\
    &\leq 4\frac{\|\ket{\delta\Psi^\beta_\alpha}\|}{\braket{\Psi_\alpha|\Psi_\alpha}} \|O\|,
\end{align}  
where we used $\braket{\Psi_\alpha|O|\Psi_\alpha}/\braket{\Psi_\alpha|\Psi_\alpha} \leq \|O\|$.  
Using Eq.~(\ref{eq:beta_norm}), this gives  
\begin{align}
    \epsilon_\beta \leq 4\|O\| e^{-\beta^2 \Delta_g^2/2} \|H'\| \Delta_g^{-1} (1-\eta)^{-1}.
\end{align}  
To ensure the projection error remains smaller than $\epsilon_\beta$, we require  
\begin{align}
    \beta \geq \frac{1}{\Delta_g} \sqrt{2 \operatorname{log}\left( \frac{\|H'\|}{\Delta_g} \frac{4\|O\|}{1-\eta} \frac{1}{\epsilon_\beta} \right)}.
\end{align}

\subsubsection{The trotterization errors}
The discussion of the Trotterization error follows a similar approach to that of the projection error. The key difference is that $\ket{\delta\Psi^\beta_\alpha}$ is replaced by $\ket{\delta\Psi^{\beta,T}_\alpha}$, and $\braket{\Psi_\alpha|\Psi_\alpha}$ is replaced by $\braket{\Psi^\beta_\alpha|\Psi^\beta_\alpha}$. Thus, the error $\epsilon_T$ in $\braket{O}$ is estimated as  
\begin{align}
    \epsilon_T \leq 4\frac{\|\ket{\delta\Psi^{\beta,T}_\alpha}\|}{\braket{\Psi^\beta_\alpha|\Psi^\beta_\alpha}} \|O\|.
\end{align}  
Using Eq.~(\ref{eq:trotter_norm}), this becomes  
\begin{align}
    \epsilon_T \leq 4 \frac{\|O\|}{1-\eta} N_\alpha \left(\frac{N_T}{2}\right)^{-p} \left(\sum_{\alpha'=1}^{N_\alpha} C_{\alpha',p}^{1/p} M^{1/p}_{1+p}(\beta)\right)^p.
\end{align}  
The Trotterization error can be kept below $\epsilon_T$
by using the total trotter steps of
\begin{align}
    N_T \geq 2 \frac{(4 N_\alpha \|O\|)^{1/p}}{(\epsilon_T (1-\eta))^{1/p}} \sum_{\alpha=1}^{N_\alpha} C_\alpha^{1/p} \braket{t^{1+p}}^{1/p}_\beta.
\end{align}  


\subsubsection{The sampling errors}
The standard error $\sigma_{\braket{O}_\alpha}$ can be estimated using a similar approach as the standard error of $E_\alpha$. This gives  
\begin{align}
    \sigma_{\braket{O}_\alpha} \leq \frac{\|O\|}{\sqrt{N_\nu}} \frac{1}{1-\eta} \sqrt{1 + \frac{1}{(1-\eta)^2}}.
\end{align}  
To achieve $\sigma_{\braket{O}_\alpha} \leq \epsilon_{\textrm{mc}}$, the number of samples $N_\nu$ can be chosen as  
\begin{align}
    N_\nu \geq \epsilon_{\textrm{mc}}^{-2} \|O\|^2 (1-\eta)^{-2} \left(1 + (1-\eta)^2\right).
\end{align}


\subsection{Remarks}
\subsubsection{Errors from inaccurate energies}
In this discussion, we assume $\Delta_g\beta \geq 1$. This assumption is reasonable because the calculation becomes infeasible otherwise.  
If there is an energy error $\epsilon$ in $E_\alpha$, the approximate projection operator is expressed as  
\begin{align}
    P^\beta(H_\alpha - E_\alpha - \epsilon) = \sum_{m_\alpha} e^{-\beta^2 (\mathcal{E}_{m,\alpha} - E_\alpha - \epsilon)^2} \ket{m_\alpha} \bra{m_\alpha}.
\end{align}  
If $\epsilon$ is comparable to $\Delta_g$, the approximate projection operator can project onto states other than the target state.  
For $\epsilon \ll \Delta_g$, we analyze the difference:  
\begin{align}
    P^\beta(H_\alpha - E_\alpha - \epsilon) - e^{-\beta^2 \epsilon^2 / 2} P^\beta(H_\alpha - E_\alpha) \nonumber \\
    = \sum_{m_\alpha \neq \Phi_\alpha} e^{-\beta^2 ((\mathcal{E}_{m,\alpha} - E_\alpha)^2 + \epsilon^2)} \nonumber \\
    \left[e^{-\beta^2 \epsilon (\mathcal{E}_{m,\alpha} - E_\alpha)} - 1\right] \ket{m_\alpha} \bra{m_\alpha}.
\end{align}  
When $\Delta_g \beta \sim 1$, the term $[\exp(-\beta^2 \epsilon (\mathcal{E}_{m,\alpha} - E_\alpha)) - 1]$ becomes small because $\epsilon \ll \Delta_g$.  
When $\Delta_g \beta \gg 1$, the factor $\exp(-\beta^2 ((\mathcal{E}_{m,\alpha} - E_\alpha)^2 + \epsilon^2))$ becomes small. 
In either case, for $\epsilon \ll \Delta_g$,
\begin{align}
    P^\beta(H_\alpha - E_\alpha - \epsilon) \approx e^{-\beta^2 \epsilon^2 / 2} P^\beta(H_\alpha - E_\alpha). \label{eq:inaccurate_energy_projection}
\end{align}  
Consecutive application of Eq.~(\ref{eq:inaccurate_energy_projection}) instead of $P^\beta(H_\alpha-E_\alpha)$
changes $\ket{\Psi^\beta_\alpha}$ to
\begin{align}
    e^{-\alpha \beta^2 \epsilon^2 / 2} \ket{\Psi_\alpha^\beta}.
\end{align}


\subsubsection{Quantum circuit implementation\label{appendix:QC}}
This section illustrates several quantum circuits employed in the Quantum Zeno Monte Carlo (QZMC) method. Figure~\ref{fig:S1} presents the QZMC circuits that uses controlled time-evolution. If the noise levels experienced by circuits (a) and (c) differ significantly, circuit (d) can be used
instead of (a)
to calculate $\braket{\Psi|\Psi}$, mitigating the influence of noise.  

The circuits shown in Figure~\ref{fig:S1} are general and applicable to any Hamiltonian. However, they can be highly susceptible to device noise due to the large number of controlled time-evolution operations. To address this limitation, the circuits in Figure~\ref{fig:S2} can be used. These circuits eliminate the need for controlled time-evolution if a common eigenstate $\ket{\Phi_\text{ref}}$ exists for $H_1, \dots, H_{N_\alpha}$.
Although this requirement may seem restrictive, many practical physical and chemical systems share a common eigenstate - the vacuum.

For Green's function calculations, it is necessary to compute $g_\alpha$ as defined in Eq.~(\ref{eq:greenmc}). Using Pauli string decomposition, this computation reduces to a weighted sum of terms such as  
\begin{align}
    \langle\Phi_0| e^{-i H_1 t_{2\alpha+1}} e^{-i H_2 t_{2\alpha}} \cdots e^{-i H_\alpha t_{\alpha+2}} P_{\gamma'} \nonumber \\
    e^{-i H_\alpha t_{\alpha+1}} P_{\gamma} e^{-i H_\alpha t_{\alpha}} \cdots e^{-i H_1 t_{1}} |\Phi_0\rangle,
\end{align}  
where $P_\gamma$ represents a Pauli string. This means that Green's function calculations can be performed using the quantum circuits shown in Figure~\ref{fig:S3}.

As noted in the method section of the main text, $1$ and $2$ qubit unitary matrix can be
compressed to a circuit with a small number of gates.
Figure.~\ref{fig:S4} depict these compressed circuit.


\begin{figure*}
\begin{flushleft}
\begin{tikzpicture}
    \node[scale=0.9] {
        \begin{quantikz}
            \quad\quad  \ket{0}
            & \gate{\textsf{H}}
            & \ctrl{1}
            & \qw \ldots
            & \ctrl{1}
            & \ctrl{1}
            & \qw \ldots
            & \ctrl{1}
            & \gate{\textsf{R}_{\textsf{z}}(\phi_\alpha)}
            & \gate{\textsf{H}} &\meter{}
            \\
            \quad\quad  \ket{0^{n}}
            & \gate{\textsf{init}}
            & \gate{e^{-iH_{1}t_1}}
            & \qw \ldots
            & \gate{e^{-iH_{\alpha}t_\alpha}}
            & \gate{e^{-iH_{\alpha}t_{\alpha+1}}}
            & \qw \ldots
            & \gate{e^{-iH_{1}t_{2\alpha}}}
            & \qw
            & \qw
            &\qw
        \end{quantikz}
    };
    \node[anchor=north west] at (current bounding box.north west) {(a)};
\end{tikzpicture}

\begin{tikzpicture}
    \node[scale=0.9] {
        \begin{quantikz}
            \quad\quad  \ket{0}
            & \gate{\textsf{H}}
            & \ctrl{1}
            & \qw \ldots
            & \ctrl{1}
            & \ctrl{1}
            & \ctrl{1}
            & \qw \ldots
            & \ctrl{1}
            & \gate{\textsf{R}_{\textsf{z}}(\phi_\alpha)}
            & \gate{\textsf{H}}
            & \meter{} 
            \\
            \quad\quad  \ket{0^{n}}
            & \gate{\textsf{init}}
            & \gate{e^{-iH_{1}t_1}}
            & \qw \ldots&\gate{e^{-iH_{\alpha}t_\alpha}}
            & \gate{P_\gamma}&\gate{e^{-iH_{\alpha}t_{\alpha+1}}}
            & \qw \ldots & \gate{e^{-iH_{1}t_{2\alpha}}}
            & \qw
            & \qw
            & \qw
        \end{quantikz}
    };
    \node[anchor=north west] at (current bounding box.north west) {(b)};
\end{tikzpicture}

\begin{tikzpicture}
    \node[scale=0.9] {
        \begin{quantikz}
            \quad\quad  \ket{0}
            & \gate{\textsf{H}}
            & \ctrl{1}
            & \qw \ldots
            & \ctrl{1}
            & \ctrl{1}
            & \ctrl{1}
            & \qw \ldots
            & \ctrl{1}
            & \gate{\textsf{R}_{\textsf{z}}(\phi_\alpha)}
            & \gate{\textsf{H}}
            & \meter{}
            \\
            \quad\quad  \ket{0^{n}}
            & \gate{\textsf{init}}
            & \gate{e^{-iH_{1}t_1}}
            & \qw \ldots
            & \gate{P_\gamma}
            & \gate{e^{-iH_{\alpha}t_\alpha}}
            & \gate{e^{-iH_{\alpha}t_{\alpha+1}}}
            & \qw \ldots
            & \gate{e^{-iH_{1}t_{2\alpha}}}
            & \qw
            & \qw 
            & \qw
        \end{quantikz}
    };
    \node[anchor=north west] at (current bounding box.north west) {(c)};
\end{tikzpicture}

\begin{tikzpicture}
    \node[scale=0.9] {
        \begin{quantikz}
            \quad\quad  \ket{0}
            & \gate{\textsf{H}}
            & \ctrl{1}
            & \qw \ldots
            & \gate[2]{\textsf{Id}_{\textsf{n}}}
            & \ctrl{1}
            & \ctrl{1}
            & \qw \ldots
            & \ctrl{1}
            & \gate{\textsf{R}_{\textsf{z}}(\phi_\alpha)}
            & \gate{\textsf{H}}
            & \meter{}
            \\
            \quad\quad  \ket{0^{n}}
            & \gate{\textsf{init}}
            & \gate{e^{-iH_{1}t_1}}
            & \qw \ldots
            & 
            & \gate{e^{-iH_{\alpha}t_\alpha}}
            & \gate{e^{-iH_{\alpha}t_{\alpha+1}}}
            & \qw \ldots
            & \gate{e^{-iH_{1}t_{2\alpha}}}
            & \qw
            & \qw 
            & \qw
        \end{quantikz}
    };
    \node[anchor=north west] at (current bounding box.north west) {(d)};
\end{tikzpicture}
\end{flushleft}

\caption{\textbf{Quantum circuits for QZMC without a reference state.}  
(a) Circuit for $\braket{\Psi_\alpha|\Psi_\alpha}$.  
(b) Circuit for $\braket{\Psi_\alpha|O|\Psi_\alpha}$.  
(c) Circuit for $\braket{\Psi_\alpha\ket{\Phi_\alpha}{\bra{\Phi_\alpha}}(H_\alpha-H_{\alpha-1})|\Psi_{\alpha-1}}$.  
The noise difference between $\braket{\Psi_\alpha|\Psi_\alpha}$ and $\braket{\Psi_\alpha\ket{\Phi_\alpha}{\bra{\Phi_\alpha}}(H_\alpha-H_{\alpha-1})|\Psi_{\alpha-1}}$ can be mitigated by using circuit (d), which includes a noisy identity operation that mimics the noise effect of controlled-$P_\gamma$, instead of circuit (a).  
The circuit \fbox{\textsf{init}} refers to the transformation of $\ket{0^{n}}$ into $\ket{\Phi_0}$. 
Here, $\phi_\alpha = \sum_{\alpha'=1}^{\alpha} E_{\alpha'} (t_{\alpha'} + t_{2\alpha + 1 - \alpha'})$
}
\label{fig:S1}
\end{figure*}

\begin{figure*}
\begin{flushleft}
\begin{tikzpicture}
    \node[scale=0.9] {
        \begin{quantikz}
            \quad\quad  \ket{0} 
            & \gate{\textsf{H}} 
            & \gate[2]{\textsf{Init}}
            & \qw
            & \qw \ldots
            & \qw
            & \qw
            & \qw \ldots
            & \qw
            & \gate[2]{\textsf{Init}^\dagger}
            & \gate{\textsf{R}_{\textsf{z}}(\phi_\alpha)}
            & \gate{\textsf{H}} &\meter{}
            \\
            \quad\quad  \ket{0^{n}}
            & \qw
            &
            & \gate{e^{-iH_{1}t_1}}
            & \qw \ldots
            & \gate{e^{-iH_{\alpha}t_\alpha}}
            & \gate{e^{-iH_{\alpha}t_{\alpha+1}}}
            & \qw \ldots
            & \gate{e^{-iH_{1}t_{2\alpha}}}
            &
            & \qw 
            & \qw 
            & \qw 
        \end{quantikz}
    };
    \node[anchor=north west] at (current bounding box.north west) {(a)};
\end{tikzpicture}

\begin{tikzpicture}
    \node[scale=0.9] {
        \begin{quantikz}
            \quad\quad \ket{0}
            & \gate{\textsf{H}} 
            & \gate[2]{\textsf{Init}}
            & \qw
            & \qw \ldots
            & \qw
            & \ctrl{1}
            & \qw
            & \qw \ldots
            & \qw
            & \gate[2]{\textsf{Init}^\dagger}
            & \gate{\textsf{R}_{\textsf{z}}(\phi_\alpha)}
            & \gate{\textsf{H}} &\meter{}
            \\
            \quad\quad \ket{0^{n}}
            & \qw
            & 
            & \gate{e^{-iH_{1}t_1}}
            & \qw \ldots
            & \gate{e^{-iH_{\alpha}t_\alpha}} 
            & \gate{P_\gamma}
            & \gate{e^{-iH_{\alpha}t_{\alpha+1}}}
            & \qw \ldots
            & \gate{e^{-iH_{1}t_{2\alpha}}}
            &
            & \qw 
            & \qw 
            & \qw
        \end{quantikz}
    };
    \node[anchor=north west] at (current bounding box.north west) {(b)};
\end{tikzpicture}

\begin{tikzpicture}
    \node[scale=0.9] {
        \begin{quantikz}
            \quad\quad \ket{0}
            & \gate{\textsf{H}}
            & \gate[2]{\textsf{Init}}
            & \qw 
            & \qw \ldots
            & \ctrl{1}
            & \qw
            & \qw
            & \qw \ldots
            & \qw
            & \gate[2]{\textsf{Init}^\dagger}
            & \gate{\textsf{R}_{\textsf{z}}(\phi_\alpha)}
            & \gate{\textsf{H}}
            & \meter{}
            \\
            \quad\quad \ket{0^{n}}
            & \qw 
            & 
            & \gate{e^{-iH_{1}t_1}} 
            & \qw \ldots 
            & \gate{P_\gamma} 
            & \gate{e^{-iH_{\alpha}t_\alpha}} 
            & \gate{e^{-iH_{\alpha}t_{\alpha+1}}}
            & \qw \ldots
            & \gate{e^{-iH_{1}t_{2\alpha}}}
            &
            & \qw
            & \qw
            & \qw
        \end{quantikz}
    };
    \node[anchor=north west] at (current bounding box.north west) {(c)};
\end{tikzpicture}

\begin{tikzpicture}
    \node[scale=0.9] {
        \begin{quantikz}
            \quad\quad  \ket{0} 
            & \gate{\textsf{H}} 
            & \gate[2]{\textsf{Init}}
            & \qw
            & \qw \ldots
            & \gate[2]{\textsf{Id}_{\textsf{n}}}
            & \qw
            & \qw
            & \qw \ldots
            & \qw 
            & \gate[2]{\textsf{Init}^\dagger}
            & \gate{\textsf{R}_{\textsf{z}}(\phi_\alpha)}
            & \gate{\textsf{H}} &\meter{}
            \\
            \quad\quad  \ket{0^{n}}
            & \qw           
            &
            & \gate{e^{-iH_{1}t_1}} 
            & \qw \ldots
            & 
            &\gate{e^{-iH_{\alpha}t_\alpha}} 
            & \gate{e^{-iH_{\alpha}t_{\alpha+1}}}
            & \qw \ldots
            & \gate{e^{-iH_{1}t_{2\alpha}}}
            &
            & \qw
            & \qw
            & \qw \\
        \end{quantikz}
    };
    \node[anchor=north west] at (current bounding box.north west) {(d)};
\end{tikzpicture}
\end{flushleft}

\caption{\textbf{Quantum circuits for QZMC with a reference state.}  
(a) Circuit for $\braket{\Psi_\alpha|\Psi_\alpha}$.  
(b) Circuit for $\braket{\Psi_\alpha|O|\Psi_\alpha}$.  
(c) Circuit for $\braket{\Psi_\alpha\ket{\Phi_\alpha}{\bra{\Phi_\alpha}}(H_\alpha-H_{\alpha-1})|\Psi_{\alpha-1}}$.  
(d) Circuit for $\braket{\Psi_\alpha|\Psi_\alpha}$, including a noisy identity operation \fbox{$\textsf{Id}_{\textsf{n}}$}, which makes the noise effect on $\braket{\Psi_\alpha|\Psi_\alpha}$ and $\braket{\Psi_\alpha\ket{\Phi_\alpha}{\bra{\Phi_\alpha}}(H_\alpha-H_{\alpha-1})|\Psi_{\alpha-1}}$ similar.  
The circuit \fbox{\textsf{init}} transforms $\ket{0^{n}}$ into a reference state $\ket{\Phi_{\textrm{ref}}}$ when the control qubit is in $\ket{0}$, and into $\ket{\Phi_0}$ when the control qubit is in $\ket{1}$.  
And the phase $\phi_\alpha$ is equal to $\sum_{\alpha'=1}^{\alpha} E_{\alpha'} (t_{\alpha'} + t_{2\alpha + 1 - \alpha'})$
}
\label{fig:S2}
\end{figure*}

\begin{figure*}
\begin{flushleft}
\begin{tikzpicture}
    \node[scale=0.84] {
        \begin{quantikz}
            \quad\quad  \ket{0}
            & \gate{\textsf{H}}
            & \ctrl{1}
            & \ctrl{1}
            & \ctrl{1}
            & \ctrl{1}
            & \ctrl{1}
            & \gate{\textsf{R}_{\textsf{z}}(\tilde{\phi}_\alpha)}
            & \gate{\textsf{H}}
            & \meter{} 
            \\
            \quad\quad  \ket{0^{n}}
            & \gate{\textsf{init}}
            & \gate{e^{-iH_{1}t_1}\cdots e^{-iH_{\alpha}t_\alpha}}
            & \gate{P_{\gamma}}
            & \gate{e^{-iH_{\alpha}t_{\alpha+1}}}
            & \gate{P_{\gamma'}}
            & \gate{e^{-iH_{\alpha}t_{\alpha+2}} \cdots e^{-iH_{1}t_{2\alpha+1}}}
            & \qw
            & \qw
            & \qw
        \end{quantikz}
    };
    \node[anchor=north west] at (current bounding box.north west) {(a)};
\end{tikzpicture}

\begin{tikzpicture}
    \node[scale=0.84] {
        \begin{quantikz}
            \quad\quad \ket{0}
            & \gate{\textsf{H}} 
            & \gate[2]{\textsf{Init}}
            & \qw
            & \ctrl{1}
            & \qw 
            & \ctrl{1}
            & \qw
            & \gate[2]{\textsf{Init}^\dagger}
            & \gate{\textsf{R}_{\textsf{z}}(\tilde{\phi}_\alpha)}
            & \gate{\textsf{H}} &\meter{}
            \\
            \quad\quad \ket{0^{n}}
            & \qw
            & 
            & \gate{e^{-iH_{1}t_1}\cdots e^{-iH_{\alpha}t_\alpha}}
            & \gate{P_{\gamma}}
            & \gate{e^{-iH_{\alpha}t_{\alpha+1}}}
            & \gate{P_{\gamma'}}
            & \gate{e^{-iH_{\alpha}t_{\alpha+2}} \cdots e^{-iH_{1}t_{2\alpha+1}}}
            &
            & \qw 
            & \qw 
            & \qw
        \end{quantikz}
    };
    \node[anchor=north west] at (current bounding box.north west) {(b)};
\end{tikzpicture}
\end{flushleft}

\caption{\textbf{Quantum circuits for Green's function.}  
(a) Circuit without a reference state.  
(b) Circuit with a reference state.  
For the phase $\tilde{\phi}_\alpha$, $\tilde{\phi}_\alpha =
\phi_\alpha = \sum_{\alpha'=1}^{\alpha} E_{\alpha'} (t_{\alpha'} + t_{2\alpha+2-\alpha'}) + (E+\omega) t_{\alpha+1}$ for the real part, and $\tilde{\phi}_\alpha = \phi_\alpha - \pi/2$ for the imaginary part.  
}
\label{fig:S3}
\end{figure*}

\section{Additional data on one and two-qubit systems}\label{appendix:NISQ}  
In this section, we give additional calculational
data on one and two qubit systems considered in the main text.


\begin{figure*}
\begin{flushleft}
\begin{tikzpicture}
    \node[scale=0.72] {
        \begin{quantikz}
            \quad\quad  \ket{0} 
            & \gate{\textsf{H}} 
            & \gate{\textsf{R}_{\textsf{z}}\left(\theta_4+\frac{\theta_3+\theta_2}{2}\right)}
            & \ctrl{1}
            & \qw
            & \ctrl{1}
            & \gate{\textsf{R}_{\textsf{z}}(\phi_\alpha)} 
            & \gate{\textsf{H}} 
            &\meter{}
            \\
            \quad\quad  \ket{0}
            & \gate{\textsf{init}}
            & \gate{\textsf{R}_{\textsf{z}}\left(\frac{\theta_3-\theta_2}{2}\right)}
            & \gate{\textsf{X}}
            & \gate{U\left(-\frac{\theta_1}{2},0,-\frac{\theta_2+\theta_3}{2}\right)}
            & \gate{\textsf{X}}
            & \gate{U\left(\frac{\theta_1}{2},\theta_2,0\right)}
            & \qw
            & \qw
        \end{quantikz}
    };
    \node[anchor=north west] at (current bounding box.north west) {(a)};
\end{tikzpicture}

\begin{tikzpicture}
    \node[scale=0.72] {
        \begin{quantikz}
            \quad\quad  \ket{0} & \gate{\textsf{H}} & \gate{\textsf{R}_{\textsf{z}}(\theta_1)}
            & \ctrl{1} 
            & \ctrl{2}
            & \ctrl{1}
            & \ctrl{1}
            & \ctrl{1}
            & \ctrl{1}
            & \ctrl{2} 
            & \gate{\textsf{R}_{\textsf{z}}(\phi_\alpha)} & \gate{\textsf{H}} &\meter{}
            \\
            \quad\quad  \ket{0} & \gate[2]{\textsf{init}}  & \qw
            & \gate{U(\theta_2,\theta_3,\theta_4)}
            & \qw
            & \gate[2]{\textsf{R}_{\textsf{XX}}(\theta_8)}
            & \gate[2]{\textsf{R}_{\textsf{YY}}(\theta_9)}
            & \gate[2]{\textsf{R}_{\textsf{YY}}(\theta_{10})}
            & \gate{U(\theta_{11},\theta_{12},\theta_{13})}
            & \qw
            & \qw
            & \qw
            & \qw
            \\
            \quad\quad  \ket{0} &  & \qw
            & \qw
            & \gate{U(\theta_5,\theta_6,\theta_7)}
            & 
            & 
            & 
            & \qw
            & \gate{U(\theta_{14},\theta_{15},\theta_{16})}
            & \qw
            & \qw
            & \qw
        \end{quantikz}
    };
    \node[anchor=north west] at (current bounding box.north west) {(b)};
\end{tikzpicture}
\end{flushleft}

\caption{\textbf{Quantum circuits for QZMC using 1- and 2-qubit circuit compression.}  
(a) Compressed circuit for 1-qubit systems.  
(b) Compressed circuit for 2-qubit systems.  
}

\label{fig:S4}
\end{figure*}

\subsection{One-Qubit System\label{sec:S1Q}}  

In the main text, we analyzed a one-qubit system described by the Hamiltonian  
\begin{align}
    H(\lambda) = \frac{X}{2} + (2\lambda - 1) Z,
\end{align}  
where $\lambda$ traverses from 0 to 1 via a discrete path defined by $\lambda_\alpha = \alpha / N_\alpha$. In this section, we use a noiseless simulator to evaluate the parameter dependencies of QZMC by varying $N_\alpha$ and $\beta$, as well as to assess the impact of finite Trotterization steps.  

\begin{figure*} 
\centering
\centerline{\includegraphics[width=17.2cm]{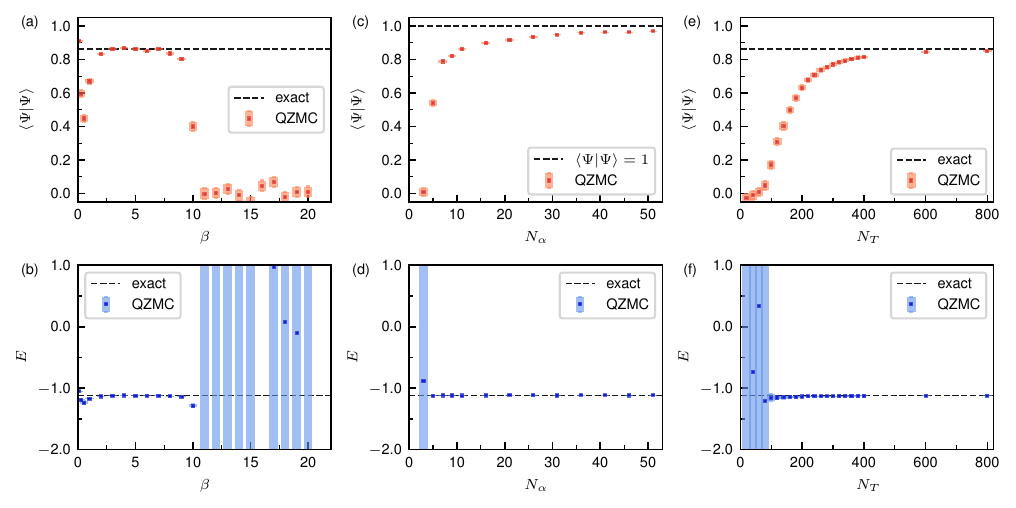}}
\caption{\textbf{Parameter dependence of QZMC applied to the one-qubit system.}  
(a) and (b) illustrate the dependence of $\braket{\Psi|\Psi}$ and $E$ on $\beta$, respectively, with $N_\alpha = 10$ and exact time evolution.  
(c) and (d) demonstrate how $\braket{\Psi|\Psi}$ and $E$ vary with $N_\alpha$, using $\beta = 5$ and exact time evolution.  
(e) and (f) plots dependence of $\braket{\Psi|\Psi}$ and $E$ on $N_T$, with $\beta = 5$ and $N_\alpha = 10$.  
Squares represent data points, while shaded regions indicate numerically estimated error bars. In this figure, we used $N_\nu = 400$.
}
\label{fig:S5}
\end{figure*}

Figure~\ref{fig:S5} illustrates these dependencies for $\braket{\Psi|\Psi}$ and the ground-state energy $E$ of $H(\lambda=1)$. Figures~\ref{fig:S5}(a) and \ref{fig:S5}(b) show the dependence on $\beta$, revealing increased accuracy with $\beta$ up to $\beta = 10$.
For $\beta > 10$, the factor $e^{-N_\alpha \beta^2 \epsilon^2}$ becomes too small,
where $\epsilon$ refers to the error in the energy estimation,
causing a significant reduction in $\braket{\Psi^\beta_\alpha|\Psi^\beta_\alpha}$,
as discussed in the main text.
This leads to large errors in the estimated energies, as seen in Fig.~\ref{fig:S5}(b).  

Figures~\ref{fig:S5}(c) and \ref{fig:S5}(d) demonstrate the variation of $\braket{\Psi|\Psi}$ and $E$ with $N_\alpha$, showing convergence to the exact values as $N_\alpha$ increases. Notably, $E$ achieves accuracy as soon as $\braket{\Psi|\Psi}$ takes on a non-trivial value, around $N_\alpha = 5$.  

Finally, Figures~\ref{fig:S5}(e) and \ref{fig:S5}(f) present results with Trotterization.
The Trotterized time evolution $U_\alpha(\tau, N_{T,\alpha})$ for $H_\alpha$ is defined as  
\begin{align}
    U_\alpha(\tau, {N_{T,\alpha}}) = \prod_{l=1}^{N_{T,\alpha}} e^{-i\Delta \tau X / 2} e^{-i\Delta \tau (2\lambda_\alpha - 1) Z},
\end{align}  
where $\Delta \tau = \tau /N_{T,\alpha}$. 
We used the same $N_{T,\alpha}$ for all $\alpha$, and the total number of Trotter steps $N_T$ was calculated as $N_T = 2\sum_{\alpha=1}^{N_\alpha} N_{T,\alpha}$.
For instance, $N_T=200$ corresponds to $N_{T,\alpha} = 10$ for each $\alpha$.

With increasing $N_T$, $\braket{\Psi|\Psi}$ converges to its exact value, while $E$ converges more rapidly. This indicates the resilience of $E$ to inaccuracies in the time evolution caused by finite Trotterization steps.


\subsection{Hubbard dimer}

We demonstrate the application of QZMC to the Hubbard dimer using Trotterized time evolution on the \textbf{\textit{ibm}}\underline{ }\textbf{\textit{perth}} quantum computer. To simplify the analysis, we focus solely on the ground state and employ second-order Trotterization for the time evolution. For $H = H_1 + H_2$, the second-order Trotterized time evolution $U(\tau, n_T)$ is defined as  
\begin{align}
    U(\tau, n_T) = \prod_{l=1}^{n_T} 
    e^{-i\Delta \tau H_1/2} e^{-i\Delta \tau H_2} e^{-i\Delta \tau H_1/2}, \label{eq:trotter2}
\end{align}  
where $\Delta \tau = \tau / n_T$.  

\begin{figure*} 
\centering
\centerline{\includegraphics[width=17.2cm]{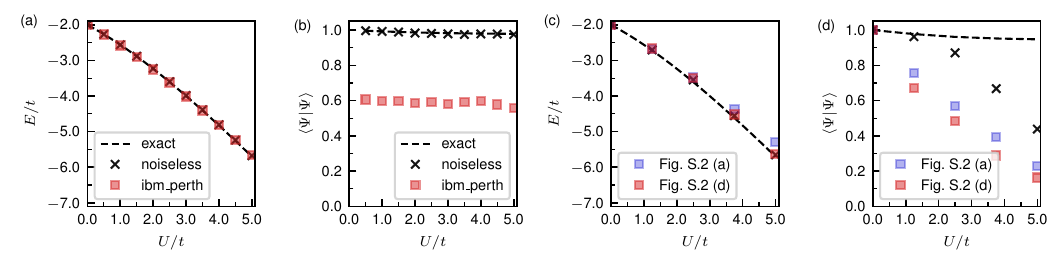}}
\caption{\textbf{QZMC for the Hubbard dimer with Trotterized time evolution.}  
(a) and (b) show results using the compressed circuit (Fig.~\ref{fig:S4}(b)) with four Trotter steps for each time evolution.  
(c) and (d) present results with uncompressed Trotterized time evolutions
with one trotter steps for each time evolution. In (c) and (d), red squares represent calculations performed with the circuit in Fig.~\ref{fig:S2}(d), while blue squares correspond to Fig.~\ref{fig:S2}(a).  
In this figure, squares indicate results obtained using \textbf{\textit{ibm}}\underline{ }\textbf{\textit{perth}}, black crosses represent noiseless simulations, and dotted lines denote the exact results. All calculations were performed with $\beta = 0.5$ and $N_\nu = 100$.  
}
\label{fig:S6}
\end{figure*}

For Figures~\ref{fig:S6}(a) and \ref{fig:S6}(b), we used the compressed circuit described in Fig.~\ref{fig:S4}(b). The Trotterization was performed using Eq.~(\ref{eq:trotter2}) with $N_{T,\alpha}=n_T = 4$ for each $\alpha$, $H_1 = -\frac{U}{2} (I + Z_1Z_2)$, and $H_2 = -t (X_1 + X_2)$. These results demonstrate the robustness of our algorithm in the presence of both Trotterization errors and device noise.  

However, circuit compression is not always feasible. To address this, we tested our algorithm with an uncompressed implementation of the Trotterization, which notably increased the circuit depth and reduced $\braket{\Psi|\Psi}$ compared to the compressed version.  

To manage the increased depth, we employed QZMC with a reference state, as depicted in Fig.~\ref{fig:S2}. The reference state used is $\ket{\Psi_\text{ref}}=[1/\sqrt{2}\quad 0\quad0\quad -1/\sqrt{2}]^T$, representing the first excited state of the Hamiltonian. Figures~\ref{fig:S6}(c) and \ref{fig:S6}(d) show results with $N_\alpha = 4$ and $N_{T,\alpha}=1$ for each $\alpha$.  

We averaged two different Trotterization choices: one with $H_2 = -\frac{U}{2} (I + Z_1Z_2)$ and the other with $H_2 = -t (X_1 + X_2)$, as the choice significantly influences computational results when the number of Trotter steps is $1$. Using the circuit in Fig.~\ref{fig:S2}(d), QZMC provided accurate energy values despite a reduction in $\braket{\Psi|\Psi}$.  

The circuits used for QZMC with uncompressed Trotterization are shown in Fig.~\ref{fig:S7}. Figure~\ref{fig:S7}(a) represents the initialization circuit, while Fig.~\ref{fig:S7}(c) mimics the noise of the controlled-\textsf{ZZ} operation depicted in Fig.~\ref{fig:S7}(b).

\begin{figure*}
\begin{minipage}{0.48\textwidth}
\begin{flushleft}
\begin{tikzpicture}
    \node[scale=0.8] {

        \begin{quantikz}[row sep = 0.3cm]
            \quad\quad
            & \gate[3]{\textsf{Init}}
            & \qw
            \\
            \quad\quad
            &\ghost{X}
            & \qw
            \\
            \quad\quad
            & 
            & 
            \qw
        \end{quantikz}=
        \begin{quantikz}[row sep = 0.3cm]
            & \ctrl{1} 
            & \ghost{X}\qw
            & \qw
            & \qw
            & \qw
            & \qw 
            \\
            & \gate{\textsf{X}}
            & \ctrl{1}   
            & \gate{\textsf{X}}
            & \gate{\textsf{H}}
            & \ctrl{1}
            & \qw 
            \\
            & \qw 
            & \gate{\textsf{H}}
            & \qw
            & \qw
            & \gate{\textsf{X}}
            & \qw 
        \end{quantikz}
    };
    \node[anchor=north west] at (current bounding box.north west) {(a)};
\end{tikzpicture}

\begin{tikzpicture}
    \node[scale=0.8] {
        \begin{quantikz}[row sep = 0.3cm]
            \quad\quad
            & \ctrl{1}
            & \ghost{X} \qw
            \\
            \quad\quad
            & \gate[2]{\textsf{ZZ}}
            & \qw
            \\
            \quad\quad
            & 
            & \qw
        \end{quantikz} =
        \begin{quantikz}[row sep = 0.3cm]
             & \qw
             & \ghost{X}\qw  
             & [0.19cm]\qw
             & \ctrl{1} 
             & [0.19cm]\qw
             & \qw 
             & \qw 
             & \qw
             \\
             & \gate{\textsf{H}}
             & \ctrl{1}
             & \qw 
             & \gate{\textsf{X}}
             & \qw 
             & \ctrl{1}
             & \gate{\textsf{H}}
             & \qw
             \\
             & \gate{\textsf{H}}
             & \gate{\textsf{X}}
             & \qw 
             & \qw 
             & \qw
             & \gate{\textsf{X}}
             & \gate{\textsf{H}}
             & \qw 
        \end{quantikz}
    };
    \node[anchor=north west] at (current bounding box.north west) {(b)};
\end{tikzpicture}

\begin{tikzpicture}
    \node[scale=0.8] {
        \begin{quantikz}[row sep = 0.3cm]
        \quad\quad
        & \gate[3]{\textsf{Id}_{\textsf{n}}}
        & \qw
        \\
        \quad\quad 
        & \ghost{X}
        & \qw
        \\
        \quad\quad 
        & 
        & \qw
        \end{quantikz} =
        \begin{quantikz}[row sep = 0.25cm]
        & \qw
        & \ghost{X}\qw
        & \gate[3]{\textsf{Delay}(t_{\textsf{CX}})}
        & \qw
        & \qw
        & \qw
        \\
        & \gate{\textsf{H}}
        & \ctrl{1}
        &
        & \ctrl{1}
        & \gate{\textsf{H}}
        & \qw 
        \\
        & \gate{\textsf{H}}
        & \gate{\textsf{X}}
        &
        & \gate{\textsf{X}}
        & \gate{\textsf{H}}
        & \qw 
        \end{quantikz}
    };
    \node[anchor=north west] at (current bounding box.north west) {(c)};
\end{tikzpicture}
\end{flushleft}

\caption {\textbf{Quantum circuits for the Hubbard dimer}.  
(a) initialization circuit for the Hubbard dimer that prepare $\ket{\Phi_0}$
if the ancilla (topmost) qubit is at $\ket{1}$ and 
$\ket{\Phi_{\textrm{ref}}}$ if the ancilla quvit is at $\ket{0}$.
(b) Implementation of the controlled-\textsf{ZZ} operation.  
(c) Noisy identity circuit used to mimic errors induced by the controlled-\textsf{ZZ} operation.  
In (c), $t_{\textsf{CX}}$ represents the duration of the CNOT operation, and Delay($t$) indicates the system waiting for a duration of $t$.  
}
\label{fig:S7}
\end{minipage}
\hfill
%
\begin{minipage}{0.48\textwidth}
    \begin{tikzpicture}
        \node[scale=0.8] {
            \begin{quantikz}[row sep = 0.3cm]
                \ket{0}
                & \qw
                & \qw
                & \qw
                & \qw
                & \gate{\textsf{H}}
                & \ctrl{1}
                & \qw
                & \qw
                & \qw
                & \qw
                \\
                \ket{0}
                & \qw
                & \qw
                & \qw
                & \gate{\textsf{X}}
                & \ctrl{-1}
                & \gate{\textsf{X}}
                & \qw
                & \qw
                & \qw
                & \qw
                \\
                \ket{0}
                & \qw
                & \qw
                & \gate{\textsf{X}}
                & \ctrl{-1}
                & \gate{\textsf{H}}
                & \gate{\textsf{Z}}
                & \ctrl{1}
                & \qw
                & \qw
                & \qw
                \\
                \ket{0}
                & \qw
                & \gate{\textsf{X}}
                & \ctrl{-1}
                & \qw
                & \ctrl{-1}
                & \qw
                & \gate{\textsf{X}}
                & \qw
                & \qw
                & \qw
                \\
                \ket{0}
                & \gate{\textsf{X}}
                & \ctrl{-1}
                & \ctrl{2}
                & \qw
                & \gate{\textsf{H}}
                & \gate{\textsf{Z}}
                & \ctrl{2}
                & \qw
                & \qw
                & \qw
                \\
                a:
                & \ctrl{-1}
                & \qw
                & \qw
                & \qw
                & \qw
                & \qw 
                & \qw
                & \qw
                & \qw
                & \qw
                \\
                \ket{0}
                & \qw
                & \qw
                & \gate{\textsf{X}}
                & \ctrl{1}
                & \ctrl{-2}
                & \qw
                & \gate{\textsf{X}}
                & \qw
                & \qw
                & \qw
                \\
                \ket{0}
                & \qw
                & \qw
                & \qw
                & \gate{\textsf{X}}
                & \ctrl{1} 
                & \qw
                & \gate{\textsf{H}}
                & \gate{\textsf{Z}}
                & \ctrl{1} 
                & \qw
                \\
                \ket{0}
                & \qw
                & \qw
                & \qw
                & \qw
                & \gate{\textsf{X}}
                & \ctrl{1}
                & \ctrl{-1}
                & \qw
                & \gate{\textsf{X}}
                & \qw
                \\
                \ket{0}
                & \qw
                & \qw
                & \qw
                & \qw
                & \qw 
                & \gate{\textsf{X}}
                & \ctrl{1}
                & \gate{\textsf{X}}
                & \qw
                & \qw
                \\
                \ket{0}
                & \qw
                & \qw
                & \qw
                & \qw
                & \qw 
                & \qw
                & \gate{\textsf{H}}
                & \ctrl{-1}
                & \qw
                & \qw
            \end{quantikz}
        };
    \end{tikzpicture}
\caption{\textbf{Quantum circuits for the XXZ model.}  
This figure shows the initialization circuit for the 10-site XXZ model. The qubit labeled "a" represents the ancilla qubit used to compute the overlap.  
}
\label{fig:S8}

\end{minipage}
\end{figure*}

\section{Quantum circuit for XXZ model}
As noted in the method section of the main text,
we start with the ground state $\ket{\Phi_0}$ of 
\begin{align}
    H_0 = -J \sum_{i;\textrm{odd}} \left(S_i^x S_{i+1}^x + S_i^y S_{i+1}^y + \Delta S_i^z S_{i+1}^z\right).
\end{align}  
The initialization circuit that prepares the vacuum state $\ket{0^n}$
when the ancilla qubit is in $\ket{0}$ and $\ket{\Phi_0}$
when the ancilla qubit is in $\ket{1}$ is depicted in Fig.~\ref{fig:S8}.


\section{Additional data on noiseless simulations}\label{appendix:noiseless}
Here, we provide additional data on the noiseless simulations in
Figures~\ref{fig:5}-\ref{fig:6} of the main text, along with the
quantum circuit in Figure~\ref{fig:S9} that implements Eq.~(\ref{eq:ground_dimer}),\
the ground state of the Hubbard dimer.

\begin{figure}
    \begin{tikzpicture}
        \node[scale=1.0] {
            \begin{quantikz}[row sep = 0.3cm]
                \fbox{$1_\uparrow$}:\, \ket{0}
                & \gate{\textsf{X}}
                & \qw
                & \qw
                & \gate{\textsf{X}}
                & \qw 
                & \qw
                \\
                \fbox{$1_\downarrow$}:\, \ket{0}
                & \gate{\textsf{X}}
                & \qw
                & \qw
                & \qw 
                & \gate{\textsf{X}}
                & \qw
                \\
                \fbox{$2_\uparrow$}:\, \ket{0}
                & \gate{\textsf{R}_\textsf{Y}(\theta_d)}
                & \ctrl{0}
                & \gate{\textsf{X}}
                & \ctrl{-2}
                & \qw 
                & \qw
                \\
                \fbox{$2_\downarrow$}:\, \ket{0}
                & \gate{\textsf{H}}
                & \ctrl{-1}
                & \ctrl{-1}
                & \qw
                & \ctrl{-2}
                & \qw
            \end{quantikz}
        };
    \end{tikzpicture}
    \caption {\textbf{Quantum circuit for the ground state of the Hubbard dimer}.  
The quantum circuit used to prepare the ground state of the Hubbard dimer is shown. The parameter $\theta_d$ is defined in Eq.~(\ref{eq:thetad}).
}
\label{fig:S9}
\end{figure}

\begin{figure*} 
\centering
\centerline{\includegraphics[width=17.2cm]{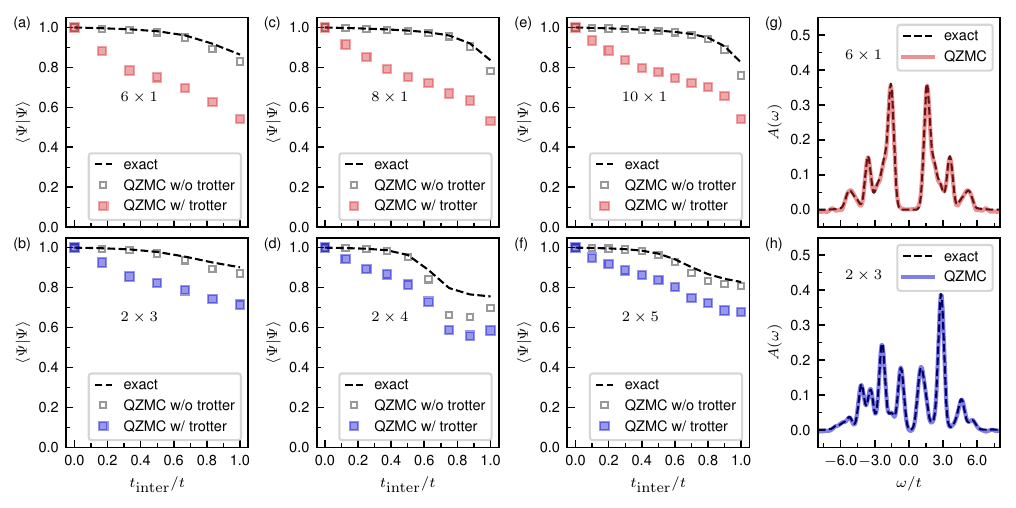}}
\caption {\textbf{The half-filled Hubbard model in various sizes with QZMC}.  
(a)-(f) show $\braket{\Psi|\Psi}$ for Hubbard models of various sizes.  
(g) and (h) present the spectral function for the 6-site Hubbard model.
}
\label{fig:S10}
\end{figure*}

Figure~\ref{fig:S10}(a)-(f) depicts $\braket{\Psi|\Psi}$
for Hubbard models studied in Fig.~\ref{fig:5}. These figures illustrate that increasing $N_\alpha$ in proportion to the number of sites yields consistent values of $\braket{\Psi|\Psi}$ across the different Hubbard models. For comparison, we also plotted $\braket{\Psi|\Psi}$ without Trotterization. All figures indicate that the Trotterization error significantly affects $\braket{\Psi|\Psi}$.  
In most cases, the Trotterization error is the dominant
source of deviation in $\braket{\Psi|\Psi}$, except for the $2 \times 4$ Hubbard model,
where the first-order perturbation energy becomes inaccurate near $\lambda = 0.7-0.8$. 
Figures~\ref{fig:S10}(g) and (h) present spectral functions calculated using QZMC for 6-site Hubbard models.

Moreover, specific values of $N_\nu$ and total Trotter steps $N_T=2 \sum_{\alpha=1}^{N_\alpha} N_{T,\alpha}$ for
each model in Fig.~\ref{fig:5} are summarized in Table~\ref{tab:numbers_hubbard}.

\begin{table} %
\caption{$N_\nu$, $N_T$, \# of gates used for Fig.~\ref{fig:5}}
\begin{tabular}{|c|c|c|c|}
\hline
Model & $N_\nu$ & $N_T$ & \# of gates\\ \hline
$6\times1$ & 576  &  528 & $1.63\times10^6$ \\ \hline
$8\times1$ & 1024 &  930 & $5.19\times10^6$ \\ \hline
$10\times1$ & 1600 & 1444 & $1.28\times10^7$ \\ \hline
$2\times3$ & 933 &  726 &  $1.97\times10^6$ \\ \hline
$2\times4$ & 1660 & 1268 & $6.02\times10^6$ \\ \hline
$2\times5$ & 2594 & 1960 & $1.45\times10^7$ \\ \hline
\end{tabular}
\label{tab:numbers_hubbard}
\end{table}

\begin{figure}
\centering
\centerline{\includegraphics[width=8.6cm]{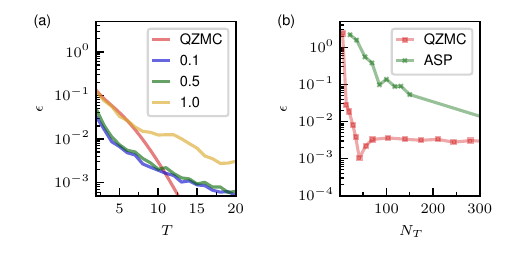}}
\caption {\textbf{Comparison of QZMC and ASP} 
The ground-state energy estimation error $\epsilon$ is
shown as a function of the maximum time length $T$
for QZMC (red) and ASP with
$\Delta t =0.1, 0.5, 1.0$ (blue, green, and yellow) in (a).
In (b), $\epsilon$ is plotted as a function of the number of Trotter steps $N_T$. The system considered is a
$4\times1$ Hubbard model at half-filling with $U/t=4$ under open boundary conditions
and the initial Hamiltonian $H_0$ is a collection of dimers.
In this figure, both ASP and QZMC start with \( \ket{\Phi_0} \), the ground state of \( H_0 \). For QZMC, we used \( N_\alpha = 4 \) and \( N_\nu = 16,384 \).
}
\label{fig:S11}
\end{figure}

Next, we give a comparison of Trotterization
dependence for our method and the adiabatic state preparation (ASP) in Fig.~\ref{fig:S11}.
For the comparison, we used a linearly interpolated Hamiltonian for ASP,
$H_j = H_0 (1-j\Delta t/T_{\textrm{ASP}}) + H j \Delta t/T_\textrm{ASP}$ and tested
several choices of $\Delta t$ in Fig.~\ref{fig:S11} (a).
Then, where QZMC and ASP (without Trotterization) yield similar accuracy
and maximum evolution times ($\Delta t =0.5$ and $T_\textrm{ASP}=9$ for ASP and $\beta=1.5$, $N_\alpha=4$ for QZMC),
we computed the ground-state energy error as a function of the number of Trotterization steps. The results clearly show that QZMC converges much faster than ASP.

\makeatletter\@input{xx.tex}\makeatother